\documentclass[jmp,aps,showpacs,reprint]{revtex4-1}
\usepackage{graphicx}
\usepackage{dcolumn}
\usepackage{amsmath}
\usepackage{xcolor}
\usepackage{epsfig}
\RequirePackage{xspace}
\usepackage{threeparttable}
\usepackage{braket}
\usepackage{relsize}
\begin{document}
	
	\title{
		\large \bfseries \boldmath Study of $B\rightarrow \rho\eta$, $\rho\eta^{\prime}$ decays in the modified perturbative QCD approach}
	\author{Yun-Han Gui} \email{guiyh@mail.nankai.edu.cn}
	\author{Mao-Zhi Yang}\email{yangmz@nankai.edu.cn}
	\affiliation{School of Physics, Nankai University, Tianjin 300071, People's Republic of China}

	\date{\today}
	
	\begin{abstract}
We study the decay processes of $B\rightarrow \rho\eta$, $\rho\eta^{\prime}$ in the perturbative quantum chromodynamics (QCD) approach with a few improvements incorporated in it, where the contributions with large momentum transfer are calculated perturbatively, and the contributions with lower energy scale are treated by introducing soft transition form factors. In addition, color-octet contributions are introduced which are essentially long-distance contributions. With reasonable input parameters selected, we find that the theoretical results of the branching ratios and $CP$ violations are all consistent with experimental data. We also predict some unmeasured quantities, which can be tested by experiment in the future.
	\end{abstract}
	\maketitle
	
	\section{Introduction}
$B$ decays are important for testing the standard model about the properties of weak interaction. It is not only weak interaction but also strong interactions are involved in $B$ decays. So it is very challenging for developing theoretical method to deal with strong interactions in $B$ decays. Several methods of treating quantum chromodynamics (QCD) effects have been developed in theory several decades before, such as perturbative QCD (PQCD) approach based on $k_t$ factorization \cite{PQCD1,PQCD2,PQCD3} and QCD factorization (QCDF) approach base on collinear factorization \cite{QCDf1,QCDf2,QCDf3,QCDf4} etc. As more and more precise data having been collected by $B$ factories \cite{PDG2024}, several deviations between theoretical predictions for branching ratios and $CP$ violations of nonleptonic decays and experimental data have been found, which are called $B\to K\pi$ and $B\to \pi\pi$ puzzles \cite{BFRS2003,LMS2005,LM2011,LM2014}. Including QCD effects of next-to-leading order can partially solve these problems and partially not in PQCD \cite{LMS2005,ZLFCX2014,Baietall2014} and QCDF \cite{BY2006,BJ2006,Bell2008,Pilipp2008,BHL2010}. There are also deviations between theory and experimental data in other nonleptonic two-body decay modes of $B$ meson \cite{C2022}. New ingredients or nonperturbative inputs are needed to diminish these deviations \cite{CC2009a,CC2009b,CCSYL2014,LLJ2016}.
	
Previously new mechanisms were developed based on the PQCD approach \cite{LY2021,LY2023,WY2023,LWY2024}, where $B$ meson wave function from relativistic potential model is used, the infrared cutoff scale $\mu_c$, the soft transition form factors and color-octet contributions are introduced. The contributions with the scale higher than the cutoff scale $\mu_c$ are calculated with PQCD approach, while the contributions with the scale lower than $\mu_c$ are considered in terms of the soft form factors. The calculations based on the modified PQCD approach can well explain the experimental data of all the $B\to PP$ decay modes \cite{LWY2024}, where $P$ stands for pseudoscalar meson.
	
In this work we study the $B\rightarrow \rho\eta$, $\rho\eta^{\prime}$ decays in this modified perturbative QCD approach. The branching ratios and $CP$ violations are calculated. With reasonable input parameters taken, the theoretical result can be well consistent with experimental measurement. We also predict some unmeasured branching ratios and $CP$ violations, which can be tested in experiment in the future.
	
The remaining part of this paper is organized as follows. The hard contributions of leading order in QCD are given in Sec. II. The perturbative contributions of next-to-leading order in QCD are presented in Sec. III. The contribution of soft transition form factors can be found in Sec. IV. Section V is for contributions of color-octet quark-antiquark pairs that finally form the final state mesons. Section VI is the numerical result and discussions. Finally Sec. VII is a brief summary.
	
	\section{The hard amplitude of leading-order in perturbative QCD}
	
	\subsection{The effective Hamiltonian}
The effective Hamiltonian of charmless hadronic weak decay of $ B $ meson caused by the $ b \rightarrow d $ transition is \cite{Hamiltanion1996}
	\begin{eqnarray}\label{Heff}
		\mathcal{H}_{\mathrm{eff}} &=& \frac{G_F}{\sqrt{2}}[V_{u}(C_{1}O_{1}^{u}+C_{2}O_{2}^{u})\nonumber\\&&-V_{t}\sum_{i=3}^{10}C_{i}O_{i}-V_{t}C_{8g}O_{8g} ],
	\end{eqnarray}
where $ V_{u}=V_{ub}V_{ud}^{*} $ and $ V_{t}=V_{tb}V_{td}^{*} $ are products of Cabibbo-Kobayashi-Maskawa (CKM) matrix elements, $G_F = 1.16638 \times 10^{-5}~\mathrm{GeV}^{-2}$ is the Fermi constant, and $C_i$'s are the Wilson coefficients. The operators in the effective Hamiltonian are given as
	\begin{eqnarray}\label{Oi}
		&&O_1^u = \bar{d}_{\alpha}\gamma^{\mu}(1-\gamma_5) u_{\beta} \bar{u}_{\beta}\gamma_{\mu}(1-\gamma_5) b_{\alpha},
		\nonumber  \\
		&&O_2^u = \bar{d}_{\alpha}\gamma^{\mu}(1-\gamma_5) u_{\alpha} \bar{u}_{\beta}\gamma_{\mu}(1-\gamma_5) b_{\beta},
		\nonumber  \\	
		&&O_3 = \bar{d}_{\alpha}\gamma^{\mu}(1-\gamma_5) b_{\alpha} \sum_{q'}\bar{q}'_{\beta}\gamma_{\mu}(1-\gamma_5) q'_{\beta},
		\nonumber  \\
		&&O_4 = \bar{d}_{\alpha}\gamma^{\mu}(1-\gamma_5) b_{\beta} \sum_{q'}\bar{q}'_{\beta}\gamma_{\mu}(1-\gamma_5) q'_{\alpha},
		\nonumber  \\	
		&&O_5 = \bar{d}_{\alpha}\gamma^{\mu}(1-\gamma_5) b_{\alpha} \sum_{q'}\bar{q}'_{\beta}\gamma_{\mu}(1+\gamma_5) q'_{\beta},
		\nonumber  \\	
		&&O_6 = \bar{d}_{\alpha}\gamma^{\mu}(1-\gamma_5) b_{\beta} \sum_{q'}\bar{q}'_{\beta}\gamma_{\mu}(1+\gamma_5) q'_{\alpha},
		\nonumber \\	
		&&O_7 = \frac{3}{2}\bar{d}_{\alpha}\gamma^{\mu}(1-\gamma_5) b_{\alpha} \sum_{q'} e_{q'} \bar{q}'_{\beta}\gamma_{\mu}(1+\gamma_5) q'_{\beta},
		\nonumber  \\	
		&&O_8 =  \frac{3}{2}\bar{d}_{\alpha}\gamma^{\mu}(1-\gamma_5) b_{\beta} \sum_{q'} e_{q'} \bar{q}'_{\beta}\gamma_{\mu}(1+\gamma_5) q'_{\alpha},
		\nonumber  \\	
		&&O_9 =  \frac{3}{2}\bar{d}_{\alpha}\gamma^{\mu}(1-\gamma_5) b_{\alpha} \sum_{q'} e_{q'} \bar{q}'_{\beta}\gamma_{\mu}(1-\gamma_5) q'_{\beta},
		\nonumber  \\	
		&&
		O_{10} =  \frac{3}{2}\bar{d}_{\alpha}\gamma^{\mu}(1-\gamma_5) b_{\beta} \sum_{q'} e_{q'} \bar{q}'_{\beta}\gamma_{\mu}(1-\gamma_5) q'_{\alpha},  	\nonumber \\
		&& O_{8\textsl{g}}=\frac{g_s}{8\pi^2}m_b\bar{d}_\alpha\sigma^{\mu\nu}(1+\gamma_5)T^a_{\alpha\beta}b_\beta G_{\mu\nu}^a,
	\end{eqnarray}
where $\alpha$ and $\beta$ are color indices, $e_{q'}$ the charge number of quark $q'$. The summation of $q'$ includes $ u,d,s,c$ and $b $ quarks.
	
	\subsection{The factorization of decay amplitude and the meson wave functions}
	
In $ B $ decays, when the momentum of the gluons exchanged between quarks exceeds the critical cutoff scale $ \mu_{c} $ which used to separate the soft and hard contributions, the decay amplitude can be factorized as the convolution of hard scattering amplitude and meson wave functions
	\begin{eqnarray}\label{factorization M}
		\mathcal{M}&=&\int d^3k \int d^3k_1 \int d^3 k_2 \Phi^B(\vec{k},\mu) C(\mu)\nonumber \\
		&&\times  H(\vec{k},\vec{k}_1,\vec{k}_2,\mu) \Phi^{\eta^{(\prime)}} (\vec{k}_1,\mu)\Phi^{\rho} (\vec{k}_2,\mu),
	\end{eqnarray}
where $\vec{k}$, $\vec{k}_1$ and $\vec{k}_2$ are three-momenta of the light quarks (or light antiquarks) in $B$, $\rho$ and $\eta^{(\prime)}$ mesons, respectively. $ H $ is the hard scattering amplitude that can be calculated by using perturbative theory, $ C(\mu) $ are the Wilson coefficients. $ \Phi^B, \Phi^{\eta^{(\prime)}}$ and $ \Phi^{\rho} $ are the meson wave functions which can be used to absorb nonperturbative soft interactions. The scale-dependence of the meson wave functions may come from the quark field renormalization for the quark and antiquarks in mesons, and the scale-dependence of the parameters in the light-cone wave functions for the fast-moving light mesons. The scale-dependence from the quark field renormalization in the meson wave functions usually appear as single ultraviolet logarithms in the exponential associated with each meson wave function, which describes the evolution of the wave function from nonperturbative scale to the hard renormalization scale. Such factor can be given associatively together with the Sudakov factor, which can be seen in Appendix A, and the Sudakov factor will be briefly explained in Sec. II C.
	
The $ B $ meson spinor wave function can be defined by the matrix element $\langle 0| \bar{q}(z)_\beta [z,0]b(0)_\alpha |\bar{B}\rangle$  as
	\begin{equation}\label{path-ordered}
		\langle 0| \bar{q}(z)_\beta [z,0]b(0)_\alpha |\bar{B}\rangle =\int d^3k \Phi^{B}_{\alpha\beta}(\vec{k},\mu)e^{-ik\cdot z},
	\end{equation}
where the four-momentum $k$ of the light quark in $B$ meson is $k^{\mu}=(E_{q},\vec{k})$ and the light quark in $B$ meson is treated to be on its mass shell with its mass $m_q\to 0$, then $E_{q}=|\vec{k}|$. $[z,0]={\cal P}\exp[-ig_sT^a\int_0^1 d\alpha z^\mu A_\mu^a(\alpha z)]$, which is the path-ordered exponential introduced to maintain the gauge invariance of the spinor wave function, $\alpha$ and $\beta$ the spinor indices of the relevant quark fields. $\Phi^{B}_{\alpha\beta}(\vec{k},\mu)$ is the spinor wave function of $B$ meson.
	
In the rest-frame of the $ B $ meson, the spinor wave function $\Phi^{B}_{\alpha\beta}(\vec{k},\mu)$ at nonperturbative scale $\mu=\Lambda_{\mathrm{QCD}}$ can be related to the wave function obtained by solving the bound-state equation within the QCD-inspired relativistic potential model \cite{Yang2012,LY2014,LY2015}. Then the spinor wave function can be derived by using Eq. (\ref{path-ordered}) and the $B$ meson wave function obtained in the potential model, which is \cite{SY2017,SY2019}
	\begin{eqnarray} \label{B-wave}
		\Phi_{\alpha\beta}^{B}(\vec{k},\Lambda_{\mathrm{QCD}})&=&\frac{-if_Bm_B}{4}K(\vec{k})
		\nonumber\\
		&& \cdot\Bigg\{(E_Q+m_Q)\frac{1+\not{v}}{2}\Bigg[\Bigg(\frac{k^+}{\sqrt{2}}  +\frac{m_q}{2}\Bigg)\not{n}_+
		\nonumber\\
		&&+\Bigg(\frac{k^-}{\sqrt{2}}  +\frac{m_q}{2}\Bigg)\not{n}_- -k_{\perp}^{\mu}\gamma_{\mu}  \Bigg]\gamma^5\nonumber\\
		&&-(E_q+m_q)\frac{1-\not{v}}{2} \Bigg[  \Bigg(\frac{k^+}{\sqrt{2}}-\frac{m_q}{2}\Bigg)\not{n}_+
		\nonumber\\
		&& +\Bigg(\frac{k^-}{\sqrt{2}}-\frac{m_q}{2}\Bigg)\not{n}_--k_{\perp}^{\mu}\gamma_{\mu}\Bigg]\gamma^5
		\Bigg\}_{\alpha\beta},\label{eqm}
	\end{eqnarray}
where $ f_{B} $ is the decay constant of $ B $ meson, $ m_{B}$ the meson mass, and $ Q $ and $ q $ represent the $ b $ and light quarks in $ B $ meson, respectively. $ v $ is the four-speed which satisfies $ v^{\mu}=(1,0,0,0) $. $ n_{\pm}^{\mu}=(1,0,0,\mp1) $ are two lightlike vectors. $ k^{\pm} $ and $ k_{\perp}^{\mu} $ are defined by
	\begin{equation}
		k^{\pm}=\dfrac{k^{0}\pm k^{3}}{\sqrt{2}} ,  k_{\perp}^{\mu}=(0,k^{1},k^{2},0),
	\end{equation}
where $ k $ is the momentum of the light quark in the rest-frame of $ B $ meson.
	
The function $K(\vec{k})$ represents a quantity involving the wave function of $ B $ meson \cite{SY2017,SY2019}
	\begin{equation}
		K(\vec{k})=\frac{2N_B\Psi_0(\vec{k})}{\sqrt{E_qE_Q(E_q+m_q)(E_Q+m_Q)}} \label{wave-k},
	\end{equation}
where $ N_{B}=\dfrac{1}{f_{B}} \sqrt{\dfrac{3}{(2\pi)^{3}m_{B}}} $ is the normalization constant. $\Psi_0(\vec{k})$ is the wave function of $ B $ meson in its rest-frame, which is obtained by numerically solving the wave equation in the QCD-inspired relativistic potential model \cite{Yang2012,LY2014,LY2015}. The numerical result for the wave function can be fitted by using the following analytical formula \cite{SY2017}
	\begin{equation}\label{psi0}
		\Psi_0(\vec{k})=a_1 e^{a_2|\vec{k}|^2+a_3|\vec{k}|+a_4},
	\end{equation}
where the obtained parameters are \cite{SY2017}
	\begin{eqnarray}
		&&a_1=4.55_{-0.30}^{+0.40}\,\mathrm{GeV}^{-3/2},\quad\;
		a_2=-0.39_{-0.20}^{+0.15}\,\mathrm{GeV}^{-2},\nonumber\\
		&& a_3=-1.55\pm 0.20\,\mathrm{GeV}^{-1},\quad   a_4=-1.10_{-0.05}^{+0.10}.
	\end{eqnarray}
	
	%\subsection{ The Mixing Scheme of $\eta-\eta' $}
For the wave functions of $\eta$ and $\eta^\prime$, the mixing scheme of $\eta$ and $\eta^\prime$ should be considered. We make use of the mixing scheme of $\eta$ and $\eta^{\prime}$ mesons suggested by Feldmann, Kroll, and Stech \cite{FKS1,FKS2}. The physical states $ \ket{\eta} $ and $ \ket{\eta^{\prime}} $ are expressed as a linear combination of orthogonal quark-flavor basis
	\begin{equation}
		\left(\begin{array}{c}
			\eta \\
			\eta^{\prime}
		\end{array}\right)=\left(\begin{array}{cc}
			\cos \phi & -\sin \phi \\
			\sin \phi & \cos \phi
		\end{array}\right)\left(\begin{array}{l}
			\eta_q \\
			\eta_s
		\end{array}\right),
	\end{equation}
where $\eta_q=(u \bar{u}+d \bar{d}) / \sqrt{2}$, $\eta_s=s \bar{s}$, and $\phi$ is the mixing angle. The decay constants for  $\eta_q$ and $\eta_s$ are defined as follows
	\begin{equation}
		\langle 0|j^{q\mu}_5|\eta_q(p)\rangle =if_q p^\mu, ~~~\langle 0|j^{s\mu}_5|\eta_s(p)\rangle =if_s p^\mu,
	\end{equation}
where $j^{q\mu}_5=\frac{\bar{u}\gamma^\mu \gamma_5 u+\bar{d}\gamma^\mu \gamma_5 d } {\sqrt{2}}$,
	$j^{s\mu}_5=\bar{s}\gamma^\mu \gamma_5 s$, and
	\begin{equation}
		\begin{array}{ll}
			\langle 0|\bar{q}\gamma^\mu\gamma_5q|\eta(p)\rangle =if_\eta^q p^\mu, &\langle 0|j^{s\mu}_5|\eta(p)\rangle =if_\eta^s p^\mu, \\
			\langle 0|\bar{q}\gamma^\mu\gamma_5q|\eta'(p)\rangle =if_{\eta'}^q p^\mu, &\langle 0|j^{s\mu}_5|\eta'(p)\rangle =if_{\eta'}^s p^\mu,
		\end{array}
	\end{equation}
here $q=u,d$. The relations between the decay constants are
	\begin{equation}
		\begin{array}{ll}
			f_\eta^q=\frac{f_q}{\sqrt{2}} \cos \phi, & f_\eta^s=-f_s \sin \phi, \\
			&  \\
			f_{\eta^{\prime}}^q=\frac{f_q}{\sqrt{2}} \sin \phi, & f_{\eta^{\prime}}^s=f_s \cos \phi.
		\end{array}
	\end{equation}
The values of the decay constants and the mixing angle are taken as \cite{FKS1,FKS2}
	\begin{equation}
		\begin{gathered}
			f_q=(1.07 \pm 0.02) f_\pi, \quad f_s=(1.34 \pm 0.06) f_\pi , \\
			\phi=39.3^{\circ} \pm 1.0^{\circ},
		\end{gathered}
	\end{equation}
where $f_\pi = 0.130 \;\mathrm{GeV}$. The chiral masses for $\eta_q$ and $\eta_s$ mesons are $\mu_{\eta_{q}}=1.07 \;\mathrm{GeV}$ and $\mu_{\eta_{s}}= 1.82\; \mathrm{GeV}$ respectively, which can be obtained by \cite{FKS1,FKS2}
	\begin{eqnarray}
		&&\mu_{\eta_{q}}=\frac{1}{2 m_q}\left(U_{11}-\frac{\sqrt{2} f_s}{f_q} U_{12}\right),\nonumber\\
		&&\mu_{\eta_{s}}=\frac{1}{2 m_s}\left(U_{22}-\frac{f_q}{\sqrt{2} f_s} U_{21}\right),
	\end{eqnarray}
where
	\begin{eqnarray}
		&&U_{11}=m_\eta^2 \cos^2 \phi  +  m_{\eta^{\prime}}^2\sin^2 \phi ,\nonumber\\
		&&U_{12}=  U_{21} = \left(m_{\eta^{\prime}}^2-m_\eta^2\right)\cos \phi \sin \phi,\nonumber\\
		&&U_{22}=  m_\eta^2 \sin^2 \phi  +  m_{\eta^{\prime}}^2\cos^2 \phi,
	\end{eqnarray}
with $m_\eta= 0.548 \;\mathrm{GeV}$ and $ m_{\eta'}=0.958 \;\mathrm{GeV}$. The masses of quarks are $m_q = 0.0056\mathrm{GeV}$ and $ m_s=0.137 \mathrm{GeV}$ \cite{Ball-Braun2006}.
	
The definition of the light-cone wave functions for the $ \eta_{q} $ and $ \eta_{s} $ is similar to that of $ \pi $ meson \cite{bra1990,bal1999}
	\begin{eqnarray}
		&&\langle \eta_{q(s)}(p_{\eta_{q(s)}})|\bar{q}(y)_\gamma q'(0)_\delta |0\rangle \nonumber\\
		&&=\int dx d^2k_{q\perp}e^{i(x p_{\eta_{q(s)}}\cdot y-y_{\perp}\cdot k_{q\perp})}\Phi^{\eta_{q(s)}}_{\delta\gamma}.
	\end{eqnarray}
In the momentum space, the spinor wave function $\Phi^{\eta_{q(s)}}_{\delta\gamma}$ can be written as \cite{bf2001,wy2002}
	\begin{eqnarray}
		\Phi^{\eta_{q(s)}}_{\delta\gamma}&=&\frac{if_{q(s)}}{4}\Bigg\{\not{p}_{\eta_{q(s)}}\gamma_5\phi_{\eta_{q(s)}}(x,k_{q\perp})
		-\mu_{\eta_{q(s)}}\gamma_5\nonumber\\
		&&\times \Bigg[\phi^{\eta_{q(s)}}_P(x,k_{q\perp}) \nonumber\\
		&&-i\sigma_{\mu\nu}\frac{p_{\eta_{q(s)}}^\mu \bar{p}_{\eta_{q(s)}}^\nu}{p_{\eta_{q(s)}}\cdot \bar{p}_{\eta_{q(s)}}} \frac{\phi'^{\eta_{q(s)}}_\sigma(x,k_{q\perp})}{6}\nonumber\\
		&&+i  \sigma_{\mu\nu}p_{\eta_{q(s)}}^\mu\frac{\phi^{\eta_{q(s)}}_\sigma(x,k_{q\perp})}{6}\frac{\partial}{\partial k_{q\perp\nu}}\Bigg]\Bigg\}_{\delta\gamma},
	\end{eqnarray}
where $ f_{q(s)} $ is the decay constant for the $\eta_{q(s)}$ meson. $ \mu_{\eta_{q(s)}} $ is the chiral mass. $\phi_{\eta_{q(s)}}$, $\phi^{\eta_{q(s)}}_P$ and $\phi^{\eta_{q(s)}}_\sigma$ are twist-2 and twist-3 distribution amplitudes, respectively. $\bar{p}_{\eta_{q(s)}} =(E_{\eta_{q(s)}},-\vec{p}_{\eta_{q(s)}})$ with $E_{\eta_{q(s)}}$ and $\vec{p}_{\eta_{q(s)}}$ are the energy and momentum of $ \eta_{q(s)} $. In addition, $\phi'^{\eta_{q(s)}}_\sigma (x,k_{q\perp})=\partial \phi^{\eta_{q(s)}}_\sigma (x,k_{q\perp})/\partial x$.
	
In $ B\rightarrow\rho\eta^{(\prime)} $ decays, the $ \rho $ meson is longitudinally polarized and its wave function in longitudinal polarization can be defined as \cite{TK-HNL,Ball-Braun1998}
	\begin{eqnarray}
		\langle \rho(p_{\rho},\epsilon_{L})|\bar{q}(y)_\gamma q'(0)_\delta |0\rangle &=&\int dx d^2k_{q\perp}e^{i(x p_\rho\cdot y-y_{\perp}\cdot k_{q\perp})}\nonumber\\
		&&\times\Phi^{\rho}_{\delta\gamma},
	\end{eqnarray}
where
	\begin{eqnarray}
		\varPhi_{\delta\gamma}^{\rho}&=&\dfrac{f_{\rho}^{||}}{4} \{ \epsilon\!\!\!/_{L} m_{\rho} \phi_{\rho}(x,k_{q\perp})+ r_{f} \epsilon\!\!\!/_{L} p\!\!\!/_{\rho} \phi_{\rho}^{t}(x,k_{q\perp})\nonumber\\
		&& + r_{f} m_{\rho} \phi_{\rho}^{s}(x,k_{q\perp}) \}_{\delta\gamma},
	\end{eqnarray}
with $ r_{f}=f_{\rho}^{\perp} /f_{\rho}^{||} $. $ f_{\rho}^{||} $ and $ f_{\rho}^{\perp} $ are the longitudinal and transverse decay constants for $ \rho $ meson, and $ m_{\rho} $ is the meson mass. $ \epsilon_{L} $ is the longitudinal polarization vector for $ \rho $ meson. $\phi_{\rho}$, $\phi_{\rho}^{t}$ and $\phi_{\rho}^{s}$ are twist-2 and twist-3 distribution amplitudes, respectively.
	%$ \epsilon\!\!\!/_{L} $
	
In principle, the wave functions for $\eta$, $\eta^\prime$ and $\rho$ mesons are also scale-dependent. But for simplicity, the explicit dependence have been omitted.
	
	\subsection{The leading order contribution}
	
Figure 1 displays eight types of diagrams at leading-order that contribute to $ B\rightarrow\rho\eta^{(\prime)} $ decays. We retain the transverse momentum of quarks and gluons in our calculations. When the transverse momentum $k_{\perp}$ and the longitudinal momentum fraction $ x $ of the quarks or gluons tends to 0, double logarithmic divergent terms such as $\alpha_s(\mu)\mbox{ln}^2 k_{\perp}/\mu$ and $\alpha_s(\mu)\mbox{ln}^2 x$ in higher-order radiative corrections of QCD will appear. The divergent terms can be resummed to Sudakov factors and the threshold factors as shown in Refs. \cite{liyu1996-1,liyu1996-2,lihn2002}. These factors can suppress the infrared contributions stemmed in the end-point regions, thereby enhancing the applicability of the perturbative calculations. For the sake of convenience in subsequent calculations, we will convert the light-cone distribution amplitudes of mesons into $ b $-space, here $b$ is the conjugate variable of the transverse momentum $k_{\perp}$, which can be found in Appendix B.
	
Figures 1(a) and 1(b) are factorizable transition diagrams, and Figs. 1(g) and 1(h) are factorizable annihilation diagrams. Figs 1(c) $\sim$ 1(f) are nonfactorizable diagrams. If $ M_{1} $ is the $\rho $ meson, Figs. 1(a) and 1(b) contribute relevant to the $ B\rightarrow\rho $ form factor. The amplitude contributed by Figs. 1(a) and 1(b) with the insertion of operators of $ (V-A)(V-A) $ is
	
	\begin{widetext}
		\begin{eqnarray}\label{Fe}
			&F_{e\rho\eta_{q(s)}}=&i2\pi^{2}\dfrac{C_{F}}{N_{c}}f_{B}f_{\rho}^{||}f_{q(s)}m_{B}^{2}\int_{\xi_{u}}^{\xi_{d}}d\xi\int_{0}^{1} dx_{1}\int_{0}^{\infty}b_{1}db_{1}b_{3}db_{3}\int dk_{\perp}k_{\perp}(\dfrac{m_{B}}{2}+\dfrac{|\vec{k_{\perp}}|^{2}}{2\xi^{2}m_{B}})K(\vec{k})(E_{Q}+m_{Q})\nonumber\\&&
			\times J_{0}(k_{\perp}b_{3})\{\alpha_{s}(t_{e}^{1})[-((x_{1}-2)E_{q}-x_{1}k^{3})\phi_{\rho}(x_{1},b_{1}) +r_{\rho}r_{f}((1-2x_{1})E_{q}+k^{3})(\phi_{\rho}^{s}(x_{1},b_{1})\nonumber\\&&
			-\phi_{\rho}^{t}(x_{1},b_{1}))]h_{e}(\xi,1-x_{1},b_{3},b_{1})S_{t}(x_{1})\exp[-S_{B}(t_{e}^{1})-S_{\rho}(t_{e}^{1})]+\alpha_{s}(t_{e}^{2})[-2r_{\rho}r_{f}(E_{q}-k^{3}) \nonumber\\&&
			\times\phi_{\rho}^{s}(x_{1},b_{1})]h_{e}(1-x_{1},\xi,b_{1},b_{3})S_{t}(\xi)\exp[-S_{B}(t_{e}^{2})-S_{\rho}(t_{e}^{2})]\},
		\end{eqnarray}
where $N_{c}=3$ is the color factor, $C_{F}=4/3$, $k_\perp=|\vec{k_{\perp}}|$, $r_{\eta_{q(s)}}=\mu_{\eta_{q(s)}}/m_{B}$,  $r_{\rho}=m_{\rho}/m_{B}$, and $\xi_{u,d}=1/2\pm\sqrt{1/4-|\vec{k_\perp}|^2/m_B^2}$ are the upper and lower integration limits for the longitudinal momentum fraction $\xi$ for the light quark in $B$ meson, respectively \cite{LY2021}. The exponentials $\exp[-S_{B,\rho,\eta_{q(s)}}(t)] $'s are the Sudakov factors associated with the relevant mesons and $ S_{t}(x) $ the threshold factor, which can be seen in Appendix A. 
		
		\begin{figure}[bth]
			\begin{center}
				\epsfig{file=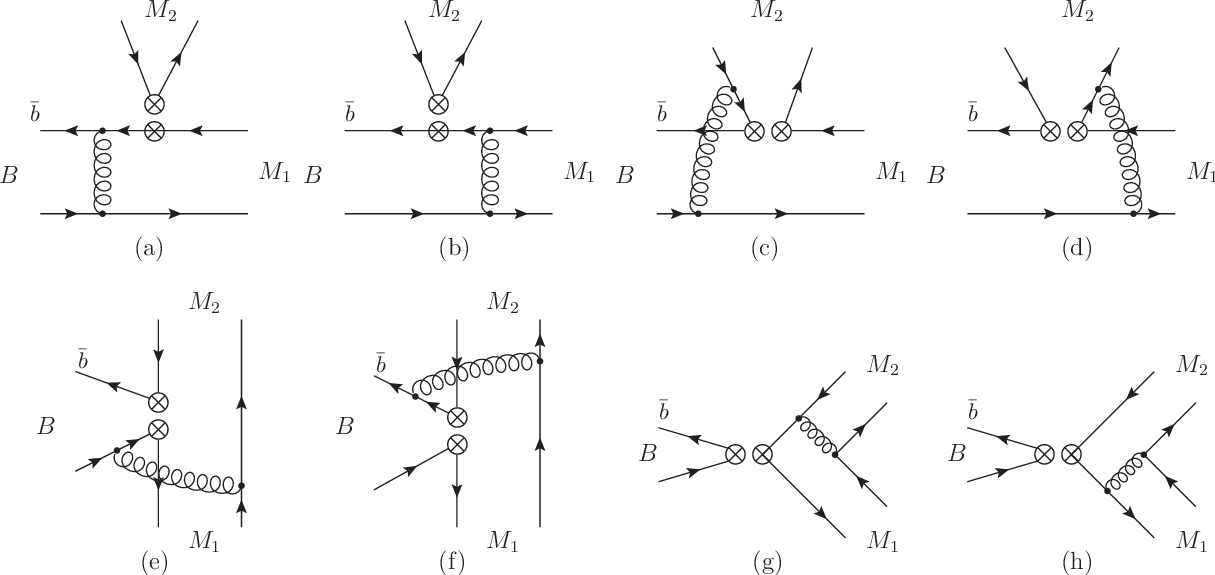,width=15cm,height=8cm}
				\caption{The Feynman diagrams of $B\rightarrow M_1 M_2$ decays at leading order, where (a) and (b) are factorizable emission diagrams, (c) and (d) nonfactorizable emission diagrams, (e) and (f) nonfactorizable annihilation diagrams, and (g), (h) factorizable annihilation diagrams.} \label{fig1}
			\end{center}
		\end{figure}
		
Due to the $ \bra{\rho}V-A\ket{B}\bra{\eta_{q(s)}}V-A\ket{0}=-\bra{\rho}V-A\ket{B}\bra{\eta_{q(s)}}V+A\ket{0} $, the contribution of $ (V-A)(V+A) $ operators  is
		\begin{eqnarray}\label{FeR}
			F_{e\rho\eta_{q(s)}}^{R}=-F_{e\rho\eta_{q(s)}}.
		\end{eqnarray}
The contribution of $(S+P)(S-P)$ operators which comes from Fierz transformation of $ (V-A)(V+A) $ operators is
		\begin{eqnarray}\label{FeP}
			&F_{e\rho\eta_{q(s)}}^{P}=&i2\pi^{2}\dfrac{C_{F}}{N_{c}}f_{B}f_{\rho}^{||}f_{q(s)}m_{B}^{2}\int_{\xi_{u}}^{\xi_{d}}d\xi\int_{0}^{1}dx_{1}\int_{0}^{\infty}b_{1}db_{1}b_{3}db_{3}\int dk_{\perp} k_{\perp}(\dfrac{m_{B}}{2}+\dfrac{|\vec{k_{\perp}}|^{2}}{2\xi^{2}m_{B}})K(\vec{k})(E_{Q}+m_{Q})\nonumber\\&&
			\times J_{0}(k_{\perp}b_{3})\{\alpha_{s}(t_{e}^{1})[-2r_{\eta_{q(s)}}(E_{q}+k^{3})\phi_{\rho}(x_{1},b_{1})-2r_{\rho}r_{\eta_{q(s)}}r_{f}((x_{1}-3)E_{q}+(1-x_{1})k^{3})\phi_{\rho}^{s}(x_{1},b_{1})\nonumber\\&&
			-2r_{\rho}r_{\eta_{q(s)}}r_{f}((x_{1}-1)E_{q}-(x_{1}+1)k^{3})\phi_{\rho}^{t}(x_{1},b_{1})]h_{e}(\xi,1-x_{1},b_{3},b_{1})S_{t}(x_{1})\exp[-S_{B}(t_{e}^{1})-S_{\rho}(t_{e}^{1})]\nonumber\\&&
			+\alpha_{s}(t_{e}^{2})[4r_{\rho}r_{\eta_{q(s)}}r_{f}(E_{q}-k^{3})\phi_{\rho}^{s}(x_{1},b_{1})]h_{e}(1-x_{1},\xi,b_{1},b_{3})S_{t}(\xi)\exp[-S_{B}(t_{e}^{2})-S_{\rho}(t_{e}^{2})]    \}.
		\end{eqnarray}
		
The contributions of Figs. 1(c) and 1(d) are
		\begin{eqnarray}\label{Me}
			&M_{e\rho\eta_{q(s)}}=&i2\pi^{2}\dfrac{C_{F}}{N_{c}}f_{B}f_{\rho}^{||}f_{q(s)}m_{B}^{2}\int_{\xi_{u}}^{\xi_{d}}d\xi\int_{0}^{1}dx_{1}dx_{2}\int_{0}^{\infty}b_{2}db_{2}b_{3}db_{3}\int dk_{\perp}k_{\perp}(\dfrac{m_{B}}{2}+\dfrac{|\vec{k_{\perp}}|^{2}}{2\xi^{2}m_{B}})K(\vec{k})(E_{Q}+m_{Q})\nonumber\\&&
			\times J_{0}(k_{\perp}b_{3})\{\alpha_{s}(t_{d}^{1})[x_{2}(E_{q}+k^{3})\phi_{\rho}(x_{1},b_{3})\phi_{\eta_{q(s)}}(x_{2},b_{2})-r_{\rho}r_{f}(x_{1}-1)(E_{q}-k^{3})(\phi_{\rho}^{s}(x_{1},b_{3})+\phi_{\rho}^{t}(x_{1},b_{3}))\nonumber\\&&
			\times\phi_{\eta_{q(s)}}(x_{2},b_{2})]h_{d}(\xi,x_{2},1-x_{1},b_{3},b_{2})\exp[-S_{B}(t_{d}^{1})-S_{\rho}(t_{d}^{1})|_{b_{1}\rightarrow b_{3}}-S_{\eta_{q(s)}}(t_{d}^{1})]+\alpha_{s}
			(t_{d}^{2})[((x_{1}+x_{2}-2)E_{q}\nonumber\\&&
			+(x_{2}-x_{1})k^{3})\phi_{\rho}(x_{1},b_{3})\phi_{\eta_{q(s)}}(x_{2},b_{2})+r_{\rho}r_{f}(x_{1}-1)(E_{q}+k^{3})(\phi_{\rho}^{s}(x_{1},b_{3})-\phi_{\rho}^{t}(x_{1},b_{3}))\phi_{\eta_{q(s)}}(x_{2},b_{2})]\nonumber\\&&
			\times h_{d}(\xi,1-x_{2},1-x_{1},b_{3},b_{2})\exp[-S_{B}(t_{d}^{2})-S_{\rho}(t_{d}^{2})|_{b_{1}\rightarrow b_{3}}-S_{\eta_{q(s)}}(t_{d}^{2})]\},
		\end{eqnarray}
		\begin{eqnarray}\label{MeR}
			&M_{e\rho\eta_{q(s)}}^{R}=&i2\pi^{2}\dfrac{C_{F}}{N_{c}}f_{B}f_{\rho}^{||}f_{q(s)}m_{B}^{2}\int_{\xi_{u}}^{\xi_{d}}d\xi\int_{0}^{1}dx_{1}dx_{2}\int_{0}^{\infty}b_{2}db_{2}b_{3}db_{3}\int dk_{\perp}k_{\perp}(\dfrac{m_{B}}{2}+\dfrac{|\vec{k_{\perp}}|^{2}}{2\xi^{2}m_{B}})K(\vec{k}) (E_{Q}+m_{Q})\nonumber\\&&
			\times J_{0}(k_{\perp}b_{3})\{\alpha_{s}(t_{d}^{1})[r_{\eta_{q(s)}}x_{2}(E_{q}+k^{3})\phi_{\rho}(x_{1},b_{3})(\dfrac{1}{6}\phi_{\sigma}^{\prime\eta_{q(s)}}(x_{2},b_{2})-\phi_{p}^{\eta_{q(s)}}(x_{2},b_{2}))+r_{\rho}r_{\eta_{q(s)}}r_{f}\nonumber\\&&
			\times((x_{1}+x_{2}-1)E_{q}-(x_{1}-x_{2}-1)k^{3})(\phi_{\rho}^{t}(x_{1},b_{3})\phi_{p}^{\eta_{q(s)}}(x_{2},b_{2})-\dfrac{1}{6}\phi_{\rho}^{s}(x_{1},b_{3})\phi_{\sigma}^{\prime\eta_{q(s)}}(x_{2},b_{2}))\nonumber\\&&
			+r_{\rho}r_{\eta_{q(s)}}r_{f}((x_{1}-x_{2}-1)E_{q}-(x_{1}+x_{2}-1)k^{3})(\dfrac{1}{6}\phi_{\rho}^{t}(x_{1},b_{3})\phi_{\sigma}^{\prime \eta_{q(s)}}(x_{2},b_{2})-\phi_{\rho}^{s}(x_{1},b_{3})\nonumber\\&&
			\times\phi_{p}^{\eta_{q(s)}}(x_{2},b_{2}))]h_{d}(\xi,x_{2},1-x_{1},b_{3},b_{2})\exp[-S_{B}(t_{d}^{1})-S_{\rho}(t_{d}^{1})|_{b_{1}\rightarrow b_{3}}-S_{\eta_{q(s)}}(t_{d}^{1})]+  \alpha_{s}(t_{d}^{2})\nonumber\\&&
			\times[-r_{\eta_{q(s)}}(x_{2}-1)(E_{q}+k^{3})\phi_{\rho}(x_{1},b_{3})(\phi_{p}^{\eta_{q(s)}}(x_{2},b_{2})+\dfrac{1}{6}\phi_{\sigma}^{\prime \eta_{q(s)}}(x_{2},b_{2}))+r_{\rho}r_{\eta_{q(s)}}r_{f}((x_{1}+x_{2}-2)E_{q}\nonumber\\&&
			+(x_{2}-x_{1})k^{3})(\phi_{\rho}^{s}(x_{1},b_{3})\phi_{p}^{\eta_{q(s)}}(x_{2},b_{2})+\dfrac{1}{6}\phi_{\rho}^{t}(x_{1},b_{3})\phi_{\sigma}^{\prime \eta_{q(s)}}(x_{2},b_{2}))+r_{\rho}r_{\eta_{q(s)}}r_{f}((x_{2}-x_{1})E_{q}\nonumber\\&&
			+(x_{1}+x_{2}-2)k^{3})(\phi_{\rho}^{t}(x_{1},b_{3})\phi_{p}^{\eta_{q(s)}}(x_{2},b_{2})+\dfrac{1}{6}\phi_{\rho}^{s}(x_{1},b_{3})\phi_{\sigma}^{\prime \eta_{q(s)}}(x_{2},b_{2}))]h_{d}(\xi,1-x_{2},1-x_{1},b_{3},b_{2})\nonumber\\&&
			\times \exp[-S_{B}(t_{d}^{2})-S_{\rho}(t_{d}^{2})|_{b_{1}\rightarrow b_{3}}-S_{\eta_{q(s)}}(t_{d}^{2})]\},
		\end{eqnarray}
		\begin{eqnarray}\label{MeP}
			&M_{e\rho\eta_{q(s)}}^{P}= &i2\pi^{2}\dfrac{C_{F}}{N_{c}}f_{B}f_{\rho}^{||}f_{q(s)}m_{B}^{2}\int_{\xi_{u}}^{\xi_{d}}d\xi\int_{0}^{1}dx_{1}dx_{2}\int_{0}^{\infty}b_{2}db_{2}b_{3}db_{3}\int dk_{\perp}k_{\perp}(\dfrac{m_{B}}{2}+\dfrac{|\vec{k_{\perp}}|^{2}}{2\xi^{2}m_{B}})K(\vec{k})(E_{Q}+m_{Q})\nonumber\\&&
			\times J_{0}(k_{\perp}b_{3})\{\alpha_{s}(t_{d}^{1})[((x_{1}-x_{2}-1)E_{q}-(x_{1}+x_{2}-1)k^{3})\phi_{\rho}(x_{1},b_{3})\phi_{\eta_{q(s)}}(x_{2},b_{2})+r_{\rho}r_{f}(x_{1}-1)\nonumber\\&&
			\times(E_{q}+k^{3})(\phi_{\rho}^{s}(x_{1},b_{3})-\phi_{\rho}^{t}(x_{1},b_{3}))\phi_{\eta_{q(s)}}(x_{2},b_{2})]h_{d}(\xi,x_{2},1-x_{1},b_{3},b_{2})\exp[-S_{B}(t_{d}^{1})-S_{\rho}(t_{d}^{1})|_{b_{1}\rightarrow b_{3}}\nonumber\\&&
			-S_{\eta_{q(s)}}(t_{d}^{1})]+\alpha_{s}(t_{d}^{2})[-(x_{2}-1)(E_{q}+k^{3})\phi_{\rho}(x_{1},b_{3})\phi_{\eta_{q(s)}}(x_{2},b_{2})-r_{\rho}r_{f}(x_{1}-1)(E_{q}-k^{3})\nonumber\\&&
			\times(\phi_{\rho}^{s}(x_{1},b_{3})+\phi_{\rho}^{t}(x_{1},b_{3}))\phi_{\eta_{q(s)}}(x_{2},b_{2})]h_{d}(\xi,1-x_{2},1-x_{1},b_{3},b_{2})\exp[-S_{B}(t_{d}^{2})-S_{\rho}(t_{d}^{2})|_{b_{1}\rightarrow b_{3}}\nonumber\\&&
			-S_{\eta_{q(s)}}(t_{d}^{2})]\}.
		\end{eqnarray}
		
The amplitudes of Figs. 1(e) and 1(f) are
		\begin{eqnarray}\label{Ma}
			&M_{a\rho\eta_{q(s)}}= &i2\pi^{2}\dfrac{C_{F}}{N_{c}}f_{B}f_{\rho}^{||}f_{q(s)}m_{B}^{2}\int_{\xi_{u}}^{\xi_{d}}d\xi\int_{0}^{1}dx_{1}dx_{2}\int_{0}^{\infty}b_{2}db_{2}b_{3}db_{3}\int dk_{\perp}k_{\perp}(\dfrac{m_{B}}{2}+\dfrac{|\vec{k_{\perp}}|^{2}}{2\xi^{2}m_{B}})K(\vec{k})(E_{Q}+m_{Q})\nonumber\\&&
			\times J_{0}(k_{\perp}b_{3})\{\alpha_{s}(t_{f}^{1})[x_{1}(E_{q}-k^{3})\phi_{\rho}(x_{1},b_{2})\phi_{\eta_{q(s)}}(x_{2},b_{2})-r_{\rho}r_{\eta_{q(s)}}r_{f}((x_{1}-x_{2}+1)E_{q}\nonumber\\&&
			-(x_{1}+x_{2}-1)k^{3})(\dfrac{1}{6}\phi_{\rho}^{t}(x_{1},b_{2})\phi_{\sigma}^{\prime\eta_{q(s)}}(x_{2},b_{2})-\phi_{\rho}^{s}(x_{1},b_{2})\phi_{p}^{\eta_{q(s)}}(x_{2},b_{2}))-r_{\rho}r_{\eta_{q(s)}}r_{f}((x_{1}+x_{2}-1)E_{q}\nonumber\\&&
			-(x_{1}-x_{2}+1)k^{3})(\dfrac{1}{6}\phi_{\rho}^{s}(x_{1},b_{2})\phi_{\sigma}^{\prime\eta_{q(s)}}(x_{2},b_{2})-\phi_{\rho}^{t}(x_{1},b_{2})\phi_{p}^{\eta_{q(s)}}(x_{2},b_{2}))]h_{f}^{1}(1-x_{2},x_{1},b_{3},b_{2})\nonumber\\&&
			\times \exp[-S_{B}(t_{f}^{1})-S_{\rho}(t_{f}^{1})|_{b_{1}\rightarrow b_{2}}-S_{\eta_{q(s)}}(t_{f}^{1})]+\alpha_{s}(t_{f}^{2})[(x_{2}-1)(E_{q}+k^{3})\phi_{\rho}(x_{1},b_{2})\phi_{\eta_{q(s)}}(x_{2},b_{2})\nonumber\\&&
			-r_{\rho}r_{\eta_{q(s)}}r_{f}((x_{1}-x_{2}+3)E_{q}-(x_{1}+x_{2}-1)k^{3})\phi_{\rho}^{s}(x_{1},b_{2})\phi_{p}^{\eta_{q(s)}}(x_{2},b_{2})+r_{\rho}r_{\eta_{q(s)}}r_{f}((x_{1}+x_{2}-1)E_{q}\nonumber\\&&
			-(x_{1}-x_{2}-1)k^{3})\phi_{\rho}^{t}(x_{1},b_{2})\phi_{p}^{\eta_{q(s)}}(x_{2},b_{2})-\dfrac{1}{6}r_{\rho}r_{\eta_{q(s)}}r_{f}((x_{1}+x_{2}-1)E_{q}-(x_{1}-x_{2}+3)k^{3})\nonumber\\&&
			\times\phi_{\rho}^{s}(x_{1},b_{2})\phi_{\sigma}^{\prime \eta_{q(s)}}(x_{2},b_{2})+\dfrac{1}{6}r_{\rho}r_{\eta_{q(s)}}r_{f}((x_{1}-x_{2}-1)E_{q}-(x_{1}+x_{2}-1)k^{3})\phi_{\rho}^{t}(x_{1},b_{2})\phi_{\sigma}^{\prime\eta_{q(s)}}(x_{2},b_{2})]\nonumber\\&&
			\times h_{f}^{2}(1-x_{2},x_{1},b_{3},b_{2})\exp[-S_{B}(t_{f}^{2})-S_{\rho}(t_{f}^{2})|_{b_{1}\rightarrow b_{2}}-S_{\eta_{q(s)}}(t_{f}^{2})]\},
		\end{eqnarray}
		\begin{eqnarray}\label{MaR}
			&M_{a\rho\eta_{q(s)}}^{R}=&i2\pi^{2}\dfrac{C_{F}}{N_{c}}f_{B}f_{\rho}^{||}f_{q(s)}m_{B}^{2}\int_{\xi_{u}}^{\xi_{d}}d\xi\int_{0}^{1}dx_{1}dx_{2}\int_{0}^{\infty}b_{2}db_{2}b_{3}db_{3}\int dk_{\perp}k_{\perp}(\dfrac{m_{B}}{2}+\dfrac{|\vec{k_{\perp}}|^{2}}{2\xi^{2}m_{B}})K(\vec{k})(E_{Q}+m_{Q})\nonumber\\&&
			\times J_{0}(k_{\perp}b_{3})\{\alpha_{s}(t_{f}^{1})[r_{\rho}r_{f}x_{1}(E_{q}+k_{3})(\phi_{\rho}^{s}(x_{1},b_{2})+\phi_{\rho}^{t}(x_{1},b_{2}))\phi_{\eta_{q(s)}}(x_{2},b_{2})+r_{\eta_{q(s)}}(x_{2}-1)(E_{q}-k^{3})\nonumber\\&&
			\times\phi_{\rho}(x_{1},b_{2})(\phi_{p}^{\eta_{q(s)}}(x_{2},b_{2})-\dfrac{1}{6}\phi_{\sigma}^{\prime\eta_{q(s)}}(x_{2},b_{2}))]h_{f}^{1}(1-x_{2},x_{1},b_{3},b_{2})\exp[-S_{B}(t_{f}^{1})-S_{\rho}(t_{f}^{1})|_{b_{1}\rightarrow b_{2}}\nonumber\\&&
			-S_{\eta_{q(s)}}(t_{f}^{1})]+\alpha_{s}(t_{f}^{2})[-r_{\rho}r_{f}((x_{1}-2)E_{q}-x_{1}k^{3})(\phi_{\rho}^{s}(x_{1},b_{2})+\phi_{\rho}^{t}(x_{1},b_{2}))\phi_{\eta_{q(s)}}(x_{2},b_{2}) \nonumber\\&&
			-r_{\eta_{q(s)}}((x_{2}+1)E_{q}+(x_{2}-1)k^{3})\phi_{\rho}(x_{1},b_{2})(\phi_{p}^{\eta_{q(s)}}(x_{2},b_{2})-\dfrac{1}{6}\phi_{\sigma}^{\prime \eta_{q(s)}}(x_{2},b_{2}))]h_{f}^{2}(1-x_{2},x_{1},b_{3},b_{2})\nonumber\\&&
			\times \exp[-S_{B}(t_{f}^{2})-S_{\rho}(t_{f}^{2})|_{b_{1}\rightarrow b_{2}}-S_{\eta_{q(s)}}(t_{f}^{2})]\},
		\end{eqnarray}
		\begin{eqnarray}\label{MaP}
			&M_{a\rho\eta_{q(s)}}^{P}= &i2\pi^{2}\dfrac{C_{F}}{N_{c}}f_{B}f_{\rho}^{||}f_{q(s)}m_{B}^{2}\int_{\xi_{u}}^{\xi_{d}}d\xi\int_{0}^{1}dx_{1}dx_{2}\int_{0}^{\infty}b_{2}db_{2}b_{3}db_{3}\int dk_{\perp}k_{\perp}(\dfrac{m_{B}}{2}+\dfrac{|\vec{k_{\perp}}|^{2}}{2\xi^{2}m_{B}})K(\vec{k})(E_{Q}+m_{Q})\nonumber\\&&
			\times J_{0}(k_{\perp}b_{3})\{\alpha_{s}(t_{f}^{1})[(x_{2}-1)(E_{q}+k^{3})\phi_{\rho}(x_{1},b_{2})\phi_{\eta_{q(s)}}(x_{2},b_{2})+r_{\rho}r_{\eta_{q(s)}}r_{f}((x_{1}+x_{2}-1)E_{q}\nonumber\\&&
			-(x_{1}-x_{2}+1)k^{3})(\phi_{\rho}^{t}(x_{1},b_{2})\phi_{p}^{\eta_{q(s)}}(x_{2},b_{2})-\dfrac{1}{6}\phi_{\rho}^{s}(x_{1},b_{2})\phi_{\sigma}^{\prime \eta_{q(s)}}(x_{2},b_{2})) +r_{\rho}r_{\eta_{q(s)}}r_{f}((x_{1}-x_{2}+1)E_{q}\nonumber\\&&
			-(x_{1}+x_{2}-1)k^{3})(\dfrac{1}{6}\phi_{\rho}^{t}(x_{1},b_{2})\phi_{\sigma}^{\prime \eta_{q(s)}}(x_{2},b_{2})-\phi_{\rho}^{s}(x_{1},b_{2})\phi_{p}^{\eta_{q(s)}}(x_{2},b_{2}))]h_{f}^{1}(1-x_{2},x_{1},b_{3},b_{2})\nonumber\\&&
			\times \exp[-S_{B}(t_{f}^{1})-S_{\rho}(t_{f}^{1})|_{b_{1}\rightarrow b_{2}}-S_{\eta_{q(s)}}(t_{f}^{1})]+\alpha_{s}(t_{f}^{2})[x_{1}(E_{q}-k^{3})\phi_{\rho}(x_{1},b_{2}) \phi_{\eta_{q(s)}}(x_{2},b_{2})\nonumber\\&&
			+r_{\rho}r_{\eta_{q(s)}}r_{f}((x_{1}-x_{2}+3)E_{q}-(x_{1}+x_{2}-1)k^{3})\phi_{\rho}^{s}(x_{1},b_{2})\phi_{p}^{\eta_{q(s)}}(x_{2},b_{2})+r_{\rho}r_{\eta_{q(s)}}r_{f}((x_{1}+x_{2}-1)E_{q}\nonumber\\&&
			-(x_{1}-x_{2}+3)k^{3})\phi_{\rho}^{t}(x_{1},b_{2})\phi_{p}^{\eta_{q(s)}}(x_{2},b_{2})-\dfrac{1}{6}r_{\rho}r_{\eta_{q(s)}}r_{f}((x_{1}+x_{2}-1)E_{q}-(x_{1}-x_{2}-1)k^{3})\phi_{\rho}^{s}(x_{1},b_{2})\nonumber\\&&
			\times \phi_{\sigma}^{\prime \eta_{q(s)}}(x_{2},b_{2})-\dfrac{1}{6}r_{\rho}r_{\eta_{q(s)}}r_{f}((x_{1}-x_{2}-1)E_{q}-(x_{1}+x_{2}-1)k^{3})  \phi_{\rho}^{t}(x_{1},b_{2})\phi_{\sigma}^{\prime \eta_{q(s)}}(x_{2},b_{2}) ] \nonumber\\&&
			\times h_{f}^{2}(1-x_{2},x_{1},b_{3},b_{2})\exp[-S_{B}(t_{f}^{2})-S_{\rho}(t_{f}^{2})|_{b_{1}\rightarrow b_{2}}-S_{\eta_{q(s)}}(t_{f}^{2})]\}.
		\end{eqnarray}
		
If $ M_{1} $ is $ \eta^{(\prime)} $ meson, then Figs. 1(a) and 1(b) contribute to $ B\rightarrow\eta_{q(s)}$ form factor. The corresponding amplitude contributed by Figs. 1(a) and 1(b) with the insertion of operators of $ (V-A)(V-A)$ is
		\begin{eqnarray}\label{Fe-eta}
			&F_{e\eta_{q(s)}\rho}= &i2\pi^{2}\dfrac{C_{F}}{N_{c}}f_{B}f_{\rho}^{||}f_{q(s)}m_{B}^{2}\int_{\xi_{u}}^{\xi_{d}}d\xi\int_{0}^{1}dx_{1}\int_{0}^{\infty}b_{1}db_{1}b_{3}db_{3}\int dk_{\perp}k_{\perp}(\dfrac{m_{B}}{2}+\dfrac{|\vec{k_{\perp}}|^{2}}{2\xi^{2}m_{B}})K(\vec{k})(E_{Q}+m_{Q})\nonumber\\&&\times
			J_{0}(k_{\perp}b_{3})\{\alpha_{s}(t_{e}^{1})[-((x_{1}-2)E_{q}-x_{1}k^{3})\phi_{\eta_{q(s)}}(x_{1},b_{1}) +r_{\eta_{q(s)}}((1-2x_{1})E_{q}+k^{3})(\dfrac{1}{6} \phi_{\sigma}^{\prime \eta_{q(s)}}(x_{1},b_{1})\nonumber\\&&
			-\phi_{p}^{\eta_{q(s)}}(x_{1},b_{1}))]h_{e}(\xi,1-x_{1},b_{3},b_{1})S_{t}(x_{1})\exp[-S_{B}(t_{e}^{1})-S_{\eta_{q(s)}}(t_{e}^{1})]+\alpha_{s}(t_{e}^{2})[2r_{\eta_{q(s)}}(E_{q}-k^{3})\nonumber\\&&
			\times \phi_{p}^{\eta_{q(s)}}(x_{1},b_{1})]h_{e}(1-x_{1},\xi,b_{1},b_{3})S_{t}(\xi)\exp[-S_{B}(t_{e}^{2})-S_{\eta_{q(s)}}(t_{e}^{2})]\},
		\end{eqnarray}
Due to the relation $ \bra{\eta_{q(s)}}V-A\ket{B}\bra{\rho}V-A\ket{0}=\bra{\eta_{q(s)}}V-A\ket{B}\bra{\rho}V+A\ket{0}$, the contribution of $ (V-A)(V+A) $ operators  is
		\begin{eqnarray}\label{FeR-eta}
			F_{e\eta_{q(s)}\rho}^{R}=F_{e\eta_{q(s)}\rho}.
		\end{eqnarray}
The contribution of $ (S+P)(S-P) $ operators is $ F_{e\eta_{q(s)\rho}}^{P}=0 $  on account of $ \bra{\rho}S+P\ket{0}=0 $.
		
Similarly, the contributions of Figs. 1(c) and 1(d) are
		\begin{eqnarray}\label{Me-eta}
			&M_{e\eta_{q(s)}\rho}=&i2\pi^{2}\dfrac{C_{F}}{N_{c}}f_{B}f_{\rho}^{||}f_{q(s)}m_{B}^{2}\int_{\xi_{u}}^{\xi_{d}}d\xi\int_{0}^{1}dx_{1} dx_{2}\int_{0}^{\infty}b_{2}db_{2}b_{3}db_{3} \int dk_{\perp}k_{\perp}(\dfrac{m_{B}}{2}+\dfrac{|\vec{k_{\perp}}|^{2}}{2\xi^{2}m_{B}})K(\vec{k})(E_{Q}+m_{Q})\nonumber\\&&
			\times J_{0}(k_{\perp}b_{3})\{\alpha_{s}(t_{d}^{1})[x_{2}(E_{q}+k^{3})\phi_{\eta_{q(s)}}(x_{1},b_{3})\phi_{\rho}(x_{2},b_{2})+r_{\eta_{q(s)}}(x_{1}-1)(E_{q}-k^{3})(\phi_{p}^{\eta_{q(s)}}(x_{1},b_{3})\nonumber\\&&
			+\dfrac{1}{6}\phi_{\sigma}^{\prime \eta_{q(s)}}(x_{1},b_{3}))\phi_{\rho}(x_{2},b_{2})] h_{d}(\xi,x_{2},1-x_{1},b_{3},b_{2})\exp[-S_{B}(t_{d}^{1})-S_{\rho}(t_{d}^{1})-S_{\eta_{q(s)}}(t_{d}^{1})|_{b_{1}\rightarrow b_{3}}] \nonumber\\&&
			+\alpha_{s}(t_{d}^{2})[((x_{1}+x_{2}-2)E_{q}+(x_{2}-x_{1})k^{3})\phi_{\eta_{q(s)}}(x_{1},b_{3})\phi_{\rho}(x_{2},b_{2})-r_{\eta_{q(s)}}(x_{1}-1)(E_{q}+k^{3})\nonumber\\&&
			\times(\phi_{p}^{\eta_{q(s)}}(x_{1},b_{3})-\dfrac{1}{6}\phi_{\sigma}^{\prime\eta_{q(s)}}(x_{1},b_{3}))\phi_{\rho}(x_{2},b_{2})]h_{d}(\xi,1-x_{2},1-x_{1},b_{3},b_{2})\nonumber\\&&
			\times \exp[-S_{B}(t_{d}^{2})-S_{\rho}(t_{d}^{2})-S_{\eta_{q(s)}}(t_{d}^{2})|_{b_{1}\rightarrow b_{3}}]\},
		\end{eqnarray}
		\begin{eqnarray}\label{MeR-eta}
			&M_{e\eta_{q(s)}\rho}^{R}=&i2\pi^{2}\dfrac{C_{F}}{N_{c}}f_{B}f_{\rho}^{||}f_{q(s)}m_{B}^{2} \int_{\xi_{u}}^{\xi_{d}}d\xi\int_{0}^{1}dx_{1} dx_{2}\int_{0}^{\infty}b_{2}db_{2}b_{3}db_{3}\int dk_{\perp} k_{\perp}(\dfrac{m_{B}}{2}+\dfrac{|\vec{k_{\perp}}|^{2}}{2\xi^{2}m_{B}}) K(\vec{k})(E_{Q}+m_{Q})\nonumber\\&&
			\times J_{0}(k_{\perp}b_{3})\{\alpha_{s}(t_{d}^{1})[-r_{\rho}r_{f}x_{2}(E_{q}+k^{3})\phi_{\eta_{q(s)}}(x_{1},b_{3})(\phi_{\rho}^{s}(x_{2},b_{2})-\phi_{\rho}^{t}(x_{2},b_{2}))+r_{\rho}r_{\eta_{q(s)}}r_{f}((x_{1}+x_{2}-1)E_{q}\nonumber\\&&
			-(x_{1}-x_{2}-1)k^{3})(\phi_{p}^{\eta_{q(s)}}(x_{1},b_{3})\phi_{\rho}^{t}(x_{2},b_{2})-\dfrac{1}{6}\phi_{\sigma}^{\prime\eta_{q(s)}}(x_{1},b_{3})\phi_{\rho}^{s}(x_{2},b_{2})) +r_{\rho}r_{\eta_{q(s)}}r_{f}((x_{1}-x_{2}-1)E_{q}\nonumber\\&&
			-(x_{1}+x_{2}-1)k^{3})(\phi_{p}^{\eta_{q(s)}}(x_{1},b_{3})\phi_{\rho}^{s}(x_{2},b_{2})-\dfrac{1}{6}\phi_{\sigma}^{\prime\eta_{q(s)}}(x_{1},b_{3})\phi_{\rho}^{t}(x_{2},b_{2}))]h_{d}(\xi,x_{2},1-x_{1},b_{3},b_{2})\nonumber\\&&
			\times \exp[-S_{B}(t_{d}^{1})-S_{\rho}(t_{d}^{1})-S_{\eta_{q(s)}}(t_{d}^{1})|_{b_{1}\rightarrow b_{3}}]+\alpha_{s}(t_{d}^{2})[-r_{\rho}r_{f}(x_{2}-1)(E_{q}+k^{3})\phi_{\eta_{q(s)}}(x_{1},b_{3})(\phi_{\rho}^{s}(x_{2},b_{2})\nonumber\\&&
			+\phi_{\rho}^{t}(x_{2},b_{2}))-r_{\rho}r_{\eta_{q(s)}}r_{f}((x_{1}+x_{2}-2)E_{q}+(x_{2}-x_{1})k^{3})(\phi_{p}^{\eta_{q(s)}}(x_{1},b_{3})\phi_{\rho}^{s}(x_{2},b_{2})\nonumber\\&&
			+\dfrac{1}{6}\phi_{\sigma}^{\prime\eta_{q(s)}}(x_{1},b_{3})\phi_{\rho}^{t}(x_{2},b_{2}))-r_{\rho}r_{\eta_{q(s)}}r_{f} ((x_{2}-x_{1})E_{q}+(x_{1}+x_{2}-2)k^{3})(\phi_{p}^{\eta_{q(s)}}(x_{1},b_{3})\phi_{\rho}^{t}(x_{2},b_{2})\nonumber\\&&
			+\dfrac{1}{6}\phi_{\sigma}^{\prime \eta_{q(s)}}(x_{1},b_{3})\phi_{\rho}^{s}(x_{2},b_{2}))]h_{d}(\xi,1-x_{2},1-x_{1},b_{3},b_{2})\exp[-S_{B}(t_{d}^{2})-S_{\rho}(t_{d}^{2})-S_{\eta_{q(s)}}(t_{d}^{2})|_{b_{1}\rightarrow b_{3}}]\},
		\end{eqnarray}
		\begin{eqnarray}\label{MeP-eta}
			&M_{e\eta_{q(s)}\rho}^{P}=&i2\pi^{2}\dfrac{C_{F}}{N_{c}}f_{B}f_{\rho}^{||}f_{q(s)}m_{B}^{2}\int_{\xi_{u}}^{\xi_{d}}d\xi\int_{0}^{1}dx_{1}dx_{2}\int_{0}^{\infty}b_{2}db_{2}b_{3}db_{3} \int dk_{\perp}k_{\perp}(\dfrac{m_{B}}{2}+\dfrac{|\vec{k_{\perp}}|^{2}}{2\xi^{2}m_{B}})K(\vec{k})(E_{Q}+m_{Q})\nonumber\\&&
			\times J_{0}(k_{\perp}b_{3})\{\alpha_{s}(t_{d}^{1})[-((x_{1}-x_{2}-1)E_{q}-(x_{1}+x_{2}-1)k^{3})\phi_{\eta_{q(s)}}(x_{1},b_{3}) \phi_{\rho}(x_{2},b_{2})-r_{\eta_{q(s)}}(x_{1}-1)\nonumber\\&&
			\times(E_{q}+k^{3})(\dfrac{1}{6}\phi_{\sigma}^{\prime \eta_{q(s)}}(x_{1},b_{3})-\phi_{p}^{\eta_{q(s)}}(x_{1},b_{3}))\phi_{\rho}(x_{2},b_{2})]h_{d}(\xi,x_{2},1-x_{1},b_{3},b_{2})\nonumber\\&&
			\times \exp[-S_{B}(t_{d}^{1})-S_{\rho}(t_{d}^{1})-S_{\eta_{q(s)}}(t_{d}^{1})|_{b_{1}\rightarrow b_{3}}]+\alpha_{s}(t_{d}^{2})[(x_{2}-1)(E_{q}+k^{3})\phi_{\eta_{q(s)}}(x_{1},b_{3})\phi_{\rho}(x_{2},b_{2})\nonumber\\&&
			-r_{\eta_{q(s)}}(x_{1}-1)(E_{q}-k^{3})(\phi_{p}^{\eta_{q(s)}}(x_{1},b_{3})+\dfrac{1}{6}\phi_{\sigma}^{\prime\eta_{q(s)}}(x_{1},b_{3}))\phi_{\rho}(x_{2},b_{2})]h_{d}(\xi,1-x_{2},1-x_{1},b_{3},b_{2}) \nonumber\\&&
			\times \exp[-S_{B}(t_{d}^{2})-S_{\rho}(t_{d}^{2})-S_{\eta_{q(s)}}(t_{d}^{2})|_{b_{1}\rightarrow b_{3}}]\}.
		\end{eqnarray}
		
The amplitudes of Figs. 1(e) and 1(f) are
		\begin{eqnarray}\label{Ma-eta}
			&M_{a\eta_{q(s)}\rho}=&i2\pi^{2}\dfrac{C_{F}}{N_{c}}f_{B}f_{\rho}^{||}f_{q(s)}m_{B}^{2}\int_{\xi_{u}}^{\xi_{d}}d\xi\int_{0}^{1}dx_{1}dx_{2}\int_{0}^{\infty}b_{2}db_{2}b_{3}db_{3} \int dk_{\perp} k_{\perp}(\dfrac{m_{B}}{2}+\dfrac{|\vec{k_{\perp}}|^{2}}{2\xi^{2}m_{B}})K(\vec{k})(E_{Q}+m_{Q})\nonumber\\&&
			\times J_{0}(k_{\perp}b_{3})\{\alpha_{s}(t_{f}^{1})[x_{1}(E_{q}-k^{3})\phi_{\eta_{q(s)}}(x_{1},b_{2})\phi_{\rho}(x_{2},b_{2})-r_{\rho}r_{\eta_{q(s)}}r_{f}((x_{1}-x_{2}+1)E_{q}\nonumber\\&&
			-(x_{1}+x_{2}-1)k^{3})(\phi_{p}^{\eta_{q(s)}}(x_{1},b_{2})\phi_{\rho}^{s}(x_{2},b_{2})-\dfrac{1}{6}\phi_{\sigma}^{\prime \eta_{q(s)}}(x_{1},b_{2})\phi_{\rho}^{t}(x_{2},b_{2})) - r_{\rho}r_{\eta_{q(s)}}r_{f} ((x_{1}+x_{2}-1)E_{q}\nonumber\\&&
			-(x_{1}-x_{2}+1)k^{3})(\dfrac{1}{6}\phi_{\sigma}^{\prime\eta_{q(s)}}(x_{1},b_{2})\phi_{\rho}^{s}(x_{2},b_{2})-\phi_{p}^{\eta_{q(s)}}(x_{1},b_{2})\phi_{\rho}^{t}(x_{2},b_{2}))]h_{f}^{1}(1-x_{2},x_{1},b_{3},b_{2})\nonumber\\&&
			\times \exp[-S_{B}(t_{f}^{1})-S_{\rho}(t_{f}^{1})-S_{\eta_{q(s)}}(t_{f}^{1})|_{b_{1}\rightarrow b_{2}}]+\alpha_{s}(t_{f}^{2})[(x_{2}-1)(E_{q}+k^{3})\phi_{\eta_{q(s)}}(x_{1},b_{2}) \phi_{\rho}(x_{2},b_{2})\nonumber\\&&
			+r_{\rho}r_{\eta_{q(s)}}r_{f}((x_{1}-x_{2}+3)E_{q}-(x_{1}+x_{2}-1)k^{3})\phi_{p}^{\eta_{q(s)}}(x_{1},b_{2})\phi_{\rho}^{s}(x_{2},b_{2})+r_{\rho}r_{\eta_{q(s)}}r_{f}((x_{1}+x_{2}-1)E_{q}\nonumber\\&&
			-(x_{1}-x_{2}+3)k^{3}) \phi_{p}^{\eta_{q(s)}}(x_{1},b_{2})\phi_{\rho}^{t}(x_{2},b_{2})-\dfrac{1}{6}r_{\rho}r_{\eta_{q(s)}}r_{f}((x_{1}+x_{2}-1)E_{q}-(x_{1}-x_{2}-1)k^{3})\nonumber\\&&
			\times\phi_{\sigma}^{\prime \eta_{q(s)}}(x_{1},b_{2})\phi_{\rho}^{s}(x_{2},b_{2})-\dfrac{1}{6}r_{\rho}r_{\eta_{q(s)}}r_{f}((x_{1}-x_{2}-1)E_{q}-(x_{1}+x_{2}-1)k^{3})\nonumber\\&&
			\times\phi_{\sigma}^{\prime \eta_{q(s)}}(x_{1},b_{2})\phi_{\rho}^{t}(x_{2},b_{2})]h_{f}^{2}(1-x_{2},x_{1},b_{3},b_{2})\exp[-S_{B}(t_{f}^{2})-S_{\rho}(t_{f}^{2})-S_{\eta_{q(s)}}(t_{f}^{2})|_{b_{1}\rightarrow b_{2}}]\},
		\end{eqnarray}
		\begin{eqnarray}\label{MaR-eta}
			&M_{a\eta_{q(s)}\rho}^{R}=&i2\pi^{2}\dfrac{C_{F}}{N_{c}}f_{B}f_{\rho}^{||}f_{q(s)}m_{B}^{2}\int_{\xi_{u}}^{\xi_{d}}d\xi\int_{0}^{1}dx_{1}dx_{2}\int_{0}^{\infty}b_{2}db_{2}b_{3}db_{3} \int dk_{\perp}k_{\perp}(\dfrac{m_{B}}{2}+\dfrac{|\vec{k_{\perp}}|^{2}}{2\xi^{2}m_{B}})K(\vec{k})(E_{Q}+m_{Q})\nonumber\\&&
			\times J_{0}(k_{\perp}b_{3})\{\alpha_{s}(t_{f}^{1})[-r_{\eta_{q(s)}}x_{1}(E_{q}+k_{3}) (\phi_{p}^{\eta_{q(s)}}(x_{1},b_{2})+\dfrac{1}{6}\phi_{\sigma}^{\prime\eta_{q(s)}}(x_{1},b_{2}))\phi_{\rho}(x_{2},b_{2})+r_{\rho}r_{f}(x_{2}-1)\nonumber\\&&
			\times(E_{q}-k^{3})\phi_{\eta_{q(s)}}(x_{1},b_{2})(\phi_{\rho}^{s}(x_{2},b_{2}) -\phi_{\rho}^{t}(x_{2},b_{2})) ]h_{f}^{1}(1-x_{2},x_{1},b_{3},b_{2})\exp[-S_{B}(t_{f}^{1})-S_{\rho}(t_{f}^{1})\nonumber\\&&
			-S_{\eta_{q(s)}}(t_{f}^{1})|_{b_{1}\rightarrow b_{2}}]+\alpha_{s}(t_{f}^{2})[r_{\eta_{q(s)}}((x_{1}-2)E_{q}-x_{1}k^{3})(\phi_{p}^{\eta_{q(s)}}(x_{1},b_{2})+\dfrac{1}{6}\phi_{\sigma}^{\prime \eta_{q(s)}}(x_{1},b_{2}))\phi_{\rho}(x_{2},b_{2}) \nonumber\\&&
			-r_{\rho}r_{f}((x_{2}+1)E_{q}+(x_{2}-1)k^{3})\phi_{\eta_{q(s)}}(x_{1},b_{2})(\phi_{\rho}^{s}(x_{2},b_{2}) -\phi_{\rho}^{t}(x_{2},b_{2}))]h_{f}^{2}(1-x_{2},x_{1},b_{3},b_{2})\nonumber\\&&
			\times \exp[-S_{B}(t_{f}^{2})-S_{\rho}(t_{f}^{2})-S_{\eta_{q(s)}}(t_{f}^{2})|_{b_{1}\rightarrow b_{2}}]\},
		\end{eqnarray}
		\begin{eqnarray}\label{MaP-eta}
			&M_{a\eta_{q(s)}\rho}^{P}= &i2\pi^{2}\dfrac{C_{F}}{N_{c}}f_{B}f_{\rho}^{||}f_{q(s)}m_{B}^{2}\int_{\xi_{u}}^{\xi_{d}}d\xi\int_{0}^{1}dx_{1}dx_{2}\int_{0}^{\infty}b_{2}db_{2}b_{3}db_{3}\int dk_{\perp} k_{\perp}(\dfrac{m_{B}}{2}+\dfrac{|\vec{k_{\perp}}|^{2}}{2\xi^{2}m_{B}})K(\vec{k})(E_{Q}+m_{Q})\nonumber\\&&
			\times  J_{0}(k_{\perp}b_{3})\{\alpha_{s}(t_{f}^{1})[(x_{2}-1)(E_{q}+k^{3})\phi_{\eta_{q(s)}}(x_{1},b_{2})\phi_{\rho}(x_{2},b_{2})+r_{\rho}r_{\eta_{q(s)}}r_{f}((x_{1}+x_{2}-1)E_{q}\nonumber\\&&
			-(x_{1}-x_{2}+1)k^{3})(\phi_{p}^{\eta_{q(s)}}(x_{1},b_{2})\phi_{\rho}^{t}(x_{2},b_{2})-\dfrac{1}{6}\phi_{\sigma}^{\prime \eta_{q(s)}}(x_{1},b_{2})\phi_{\rho}^{s}(x_{2},b_{2})) +r_{\rho}r_{\eta_{q(s)}}r_{f}((x_{1}-x_{2}+1)E_{q}\nonumber\\&&
			-(x_{1}+x_{2}-1)k^{3})(\phi_{p}^{\eta_{q(s)}}(x_{1},b_{2})\phi_{\rho}^{s}(x_{2},b_{2})-\dfrac{1}{6}\phi_{\sigma}^{\prime \eta_{q(s)}}(x_{1},b_{2})\phi_{\rho}^{t}(x_{2},b_{2}))]h_{f}^{1}(1-x_{2},x_{1},b_{3},b_{2})\nonumber\\&&
			\times \exp[-S_{B}(t_{f}^{1})-S_{\rho}(t_{f}^{1})-S_{\eta_{q(s)}}(t_{f}^{1})|_{b_{1}\rightarrow b_{2}}]+\alpha_{s}(t_{f}^{2})[x_{1}(E_{q}-k^{3})\phi_{\eta_{q(s)}}(x_{1},b_{2})\phi_{\rho}(x_{2},b_{2}) \nonumber\\&&
			-r_{\rho}r_{\eta_{q(s)}}r_{f}((x_{1}-x_{2}+3)E_{q}-(x_{1}+x_{2}-1)k^{3})\phi_{p}^{\eta_{q(s)}}(x_{1},b_{2})\phi_{\rho}^{s}(x_{2},b_{2})+r_{\rho}r_{\eta_{q(s)}}r_{f}\nonumber\\&&
			\times ((x_{1}+x_{2}-1)E_{q}-(x_{1}-x_{2}-1)k^{3})\phi_{p}^{\eta_{q(s)}}(x_{1},b_{2})\phi_{\rho}^{t}(x_{2},b_{2})-\dfrac{1}{6}r_{\rho}r_{\eta_{q(s)}}r_{f}((x_{1}+x_{2}-1)E_{q}\nonumber\\&&
			-(x_{1}-x_{2}+3)k^{3})\phi_{\sigma}^{\prime \eta_{q(s)}}(x_{1},b_{2})\phi_{\rho}^{s}(x_{2},b_{2})+\dfrac{1}{6}r_{\rho}r_{\eta_{q(s)}}r_{f}((x_{1}-x_{2}-1)E_{q}-(x_{1}+x_{2}-1)k^{3})\nonumber\\&&
			\times\phi_{\sigma}^{\prime\eta_{q(s)}}(x_{1},b_{2})\phi_{\rho}^{t}(x_{2},b_{2})]h_{f}^{2}(1-x_{2},x_{1},b_{3},b_{2})\exp[-S_{B}(t_{f}^{2})-S_{\rho}(t_{f}^{2})-S_{\eta_{q(s)}}(t_{f}^{2})|_{b_{1}\rightarrow b_{2}}]\}.
		\end{eqnarray}
Particularly, there are no contributions from Figs. 1(g) and 1(h) in the decay modes of $ B\rightarrow\rho\eta^{(\prime)}$. 
		
In Eqs. (\ref{Fe})$ - $(\ref{MaP-eta}), the functions $ h_{i}$’s are given as
		\begin{eqnarray}
			h_{e}(x_{1},x_{2},b_{1},b_{2})&=&K_{0}(\sqrt{x_{1}x_{2}}m_{B}b_{1})[\theta(b_{1}-b_{2})I_{0}(\sqrt{x_{2}}m_{B}b_{2})K_{0}(\sqrt{x_{2}}m_{B}b_{1})\nonumber\\&&
			+\theta(b_{2}-b_{1})I_{0}(\sqrt{x_{2}}m_{B}b_{1})K_{0}(\sqrt{x_{2}}m_{B}b_{2}) ],\nonumber\\&&
		\end{eqnarray}
		\begin{eqnarray}
			h_{d}(x_{1},x_{2},x_{3},b_{1},b_{2})&=&K_{0}(-i\sqrt{x_{2}x_{3}}m_{B}b_{2})[\theta(b_{1}-b_{2})I_{0}(\sqrt{x_{1}x_{3}}m_{B}b_{2})K_{0}(\sqrt{x_{1}x_{3}}m_{B}b_{1})\nonumber\\&&
			+\theta(b_{2}-b_{1})I_{0}(\sqrt{x_{1}x_{3}}m_{B}b_{1})K_{0}(\sqrt{x_{1}x_{3}}m_{B}b_{2})],\nonumber\\&&
		\end{eqnarray}
		\begin{eqnarray}
			h_{f}^{1}(x_{1},x_{2},b_{1},b_{2})&=&K_{0}(-i\sqrt{x_{1}x_{2}}m_{B}b_{1})[\theta(b_{1}-b_{2})I_{0}(-i\sqrt{x_{1}x_{2}}m_{B}b_{2})K_{0}(-i\sqrt{x_{1}x_{2}}m_{B}b_{1})\nonumber\\&&
			+\theta(b_{2}-b_{1})I_{0}(-i\sqrt{x_{1}x_{2}}m_{B}b_{1})K_{0}(-i\sqrt{x_{1}x_{2}}m_{B}b_{2}) ],\nonumber\\&&
		\end{eqnarray}
		\begin{eqnarray}
			h_{f}^{2}(x_{1},x_{2},b_{1},b_{2})&=&K_{0}(\sqrt{x_{1}+x_{2}-x_{1}x_{2}}m_{B}b_{1})[\theta(b_{1}-b_{2})I_{0}(-i\sqrt{x_{1}x_{2}}m_{B}b_{2}) K_{0}(-i\sqrt{x_{1}x_{2}}m_{B}b_{1})\nonumber\\&&+\theta(b_{2}-b_{1})I_{0}(-i\sqrt{x_{1}x_{2}}m_{B}b_{1})K_{0}(-i\sqrt{x_{1}x_{2}}m_{B}b_{2})].\nonumber\\&&
		\end{eqnarray}
		
In order to suppress the large logarithmic terms in higher-order corrections, the hard scales are taken as the maximum mass scales in the amplitudes
		\begin{eqnarray}
			t_{e}^{1}&=&max(\sqrt{1-x_{1}}m_{B},1/b_{3},1/b_{1}),\nonumber\\
			t_{e}^{2}&=&max(\sqrt{\xi}m_{B},1/b_{3},1/b_{1}),\nonumber\\
			%t_{a}^{1}&=&max(\sqrt{x_{1}}m_{B},1/b_{1},1/b_{2}),\nonumber\\
			%t_{a}^{2}&=&max(\sqrt{1-x_{2}}m_{B},1/b_{1},1/b_{2}),\nonumber\\
			t_{d}^{1}&=&max(\sqrt{(1-x_{1})x_{2}}m_{B},\sqrt{(1-x_{1})\xi}m_{B},1/b_{3},1/b_{2}),\nonumber\\
			t_{d}^{2}&=&max(\sqrt{(1-x_{1})(1-x_{2})}m_{B},\sqrt{(1-x_{1})\xi}m_{B},1/b_{3},1/b_{2}),\nonumber\\
			t_{f}^{1}&=&max(\sqrt{x_{1}(1-x_{2})}m_{B},1/b_{3},1/b_{2}),\nonumber\\
			t_{f}^{2}&=&max(\sqrt{x_{1}(1-x_{2})}m_{B},\sqrt{x_{1}+(1-x_{2})-x_{1}(1-x_{2})}m_{B},1/b_{3},1/b_{2}).
		\end{eqnarray}
		
The decay amplitudes of the $ B\rightarrow \rho\eta_{q,s} $ are
		\begin{eqnarray}\label{Mrho-etaq}
			\sqrt{2}\mathcal{M}(B^{-}\rightarrow\rho^{-}\eta_{q})&=&\{V_{u}(C_{1}+\dfrac{C_{2}}{N_{c}})-V_{t}[(2+\dfrac{1}{N_{c}})C_{3}+(1+\dfrac{2}{N_{c}})C_{4}\ +(1-\dfrac{1}{N_{c}})\dfrac{1}{2}C_{9}-(1-\dfrac{1}{N_{c}})\dfrac{1}{2}C_{10}]\}F_{e\rho\eta_{q}}\nonumber\\&&
			-V_{t}(2C_{5}+2\dfrac{C_{6}}{N_{c}}+\dfrac{1}{2}C_{7}+\dfrac{1}{2}\dfrac{C_{8}}{N_{c}})F_{e\rho\eta_{q}}^{R} -V_{t}(\dfrac{C_{5}}{N_{c}}+C_{6}-\dfrac{1}{2}\dfrac{C_{7}}{N_{c}}-\dfrac{1}{2}C_{8})F_{e\rho\eta_{q}}^{P}\nonumber\\&&
			+[V_{u}(\dfrac{C_{1}}{N_{c}}+C_{2})-V_{t}(\dfrac{C_{3}}{N_{c}}+C_{4}+\dfrac{C_{9}}{N_{c}}+C_{10})]F_{e\eta_{q}\rho} -V_{t}(\dfrac{C_{5}}{N_{c}}+C_{6}+\dfrac{C_{7}}{N_{c}}+C_{8})F_{e\eta_{q}\rho}^{P}\nonumber\\&&
			+[V_{u}\dfrac{C_{2}}{N_{c}}-V_{t}\dfrac{1}{N_{c}}(C_{3}+2C_{4}-\dfrac{1}{2}C_{9} +\dfrac{1}{2}C_{10})]M_{e\rho\eta_{q}}-V_{t}\dfrac{1}{N_{c}}(C_{5}-\dfrac{1}{2}C_{7})M_{e\rho\eta_{q}}^{R}\nonumber\\&&
			-V_{t}\dfrac{1}{N_{c}}(2C_{6}+\dfrac{1}{2}C_{8})M_{e\rho\eta_{q}}^{P}+[V_{u}\dfrac{C_{1}}{N_{c}}-V_{t}\dfrac{1}{N_{c}}(C_{3}+C_{9})](M_{e\eta_{q}\rho}+M_{a\rho\eta_{q}}+M_{a\eta_{q}\rho})\nonumber\\&&
			-V_{t}\dfrac{1}{N_{c}}(C_{5}+C_{7})(M_{e\eta_{q}\rho}^{R}+M_{a\rho\eta_{q}}^{R}+M_{a\eta_{q}\rho}^{R}),
		\end{eqnarray}
		\begin{eqnarray}\label{Mrho-etas}
			\mathcal{M}(B^{-}\rightarrow\rho^{-}\eta_{s})&=&-V_{t}(C_{3}+\dfrac{C_{4}}{N_{c}}-\dfrac{1}{2}C_{9}-\dfrac{1}{2}\dfrac{C_{10}}{N_{c}})F_{e\rho\eta_{s}}-V_{t}(C_{5}+\dfrac{C_{6}}{N_{c}}-\dfrac{1}{2}C_{7}-\dfrac{1}{2}\dfrac{C_{8}}{N_{c}})F_{e\rho\eta_{s}}^{R}\nonumber\\&&
			-V_{t}\dfrac{1}{N_{c}}(C_{4}-\dfrac{1}{2}C_{10})M_{e\rho\eta_{s}}-V_{t}\dfrac{1}{N_{c}}(C_{6}-\dfrac{1}{2}C_{8})M_{e\rho\eta_{s}}^{P}.
		\end{eqnarray}
		\begin{eqnarray}\label{Mrho0baretaq}
			2\mathcal{M}(\bar{B^{0}}\rightarrow\rho^{0}\eta_{q})&=&-\{V_{u}(C_{1}+\dfrac{C_{2}}{N_{c}})-V_{t}[(2+\dfrac{1}{N_{c}})C_{3}+(1+\dfrac{2}{N_{c}})C_{4} +(1-\dfrac{1}{N_{c}})\dfrac{1}{2}C_{9}-(1-\dfrac{1}{N_{c}})\dfrac{1}{2}C_{10}  ]  \}F_{e\rho\eta_{q}}\nonumber\\&&
			+\{V_{u}(C_{1}+\dfrac{C_{2}}{N_{c}})-V_{t}[-\dfrac{C_{3}}{N_{c}}-C_{4}+(3+\dfrac{1}{N_{c}})\dfrac{1}{2}C_{9}+(1+\dfrac{3}{N_{c}})\dfrac{1}{2}C_{10}  ]  \}F_{e\eta_{q}\rho}\nonumber\\&&
			+V_{t}(2C_{5}+2\dfrac{C_{6}}{N_{c}}+\dfrac{1}{2}C_{7}+\dfrac{1}{2}\dfrac{C_{8}}{N_{c}})F_{e\rho\eta_{q}}^{R}
			-V_{t}(\dfrac{3}{2}C_{7}+\dfrac{3}{2}\dfrac{C_{8}}{N_{c}})F_{e\eta_{q}\rho}^{R}\nonumber\\&&
			+V_{t}(\dfrac{C_{5}}{N_{c}}+C_{6}-\dfrac{1}{2}\dfrac{C_{7}}{N_{c}}-\dfrac{1}{2}C_{8})(F_{e\rho\eta_{q}}^{P}+F_{e\eta_{q}\rho}^{P})
			\nonumber\\&&
			-[V_{u}\dfrac{C_{2}}{N_{c}}-V_{t}\dfrac{1}{N_{c}}(C_{3}+2C_{4}-\dfrac{1}{2}C_{9}+\dfrac{1}{2}C_{10})]M_{e\rho\eta_{q}}\nonumber\\&&
			+[V_{u}\dfrac{C_{2}}{N_{c}}-V_{t}\dfrac{1}{N_{c}}(-C_{3}+\dfrac{1}{2}C_{9}+\dfrac{3}{2}C_{10})](M_{e\eta_{q}\rho}+M_{a\rho\eta_{q}}+M_{a\eta_{q}\rho})\nonumber\\&&
			+V_{t}\dfrac{1}{N_{c}}(C_{5}-\dfrac{1}{2}C_{7})(M_{e\rho\eta_{q}}^{R}+M_{e\eta_{q}\rho}^{R}+M_{a\rho\eta_{q}}^{R}+M_{a\eta_{q}\rho}^{R})
			+V_{t}\dfrac{1}{N_{c}}(2C_{6}+\dfrac{1}{2}C_{8})M_{e\rho\eta_{q}}^{P}\nonumber\\&&
			-V_{t}\dfrac{1}{N_{c}}\dfrac{3}{2}C_{8}(M_{e\eta_{q}\rho}^{P}+M_{a\rho\eta_{q}}^{P}+M_{a\eta_{q}\rho}^{P}),
		\end{eqnarray}
		\begin{eqnarray}\label{Mrho0baretas}
			\sqrt{2}\mathcal{M}(\bar{B^{0}}\rightarrow\rho^{0}\eta_{s})&=&
			V_{t}(C_{3}+\dfrac{C_{4}}{N_{c}}-\dfrac{1}{2}C_{9}-\dfrac{1}{2}\dfrac{C_{10}}{N_{c}})F_{e\rho\eta_{s}}
			+V_{t}(C_{5}+\dfrac{C_{6}}{N_{c}}-\dfrac{1}{2}C_{7}-\dfrac{1}{2}\dfrac{C_{8}}{N_{c}})F_{e\rho\eta_{s}}^{R}\nonumber\\&&
			+V_{t}\dfrac{1}{N_{c}}(C_{4}-\dfrac{1}{2}C_{10})M_{e\rho\eta_{s}}+V_{t}\dfrac{1}{N_{c}}(C_{6}-\dfrac{1}{2}C_{8})M_{e\rho\eta_{s}}^{P}.
		\end{eqnarray}
		
		Considering the mixing scheme of $\eta$ and  $\eta^\prime$, the decay amplitude of $B\to \rho\eta^{(\prime)}$ can be written as
		\begin{eqnarray}
			\mathcal{M}(B^{-}\rightarrow\rho^{-}\eta)=\cos\phi\cdot \mathcal{M}(B^{-}\rightarrow\rho^{-}\eta_{q}) -\sin\phi \cdot \mathcal{M}(B^{-}\rightarrow\rho^{-}\eta_{s}),
		\end{eqnarray}
		\begin{eqnarray}
			\mathcal{M}(B^{-}\rightarrow\rho^{-}\eta^{\prime})=\sin\phi\cdot \mathcal{M}(B^{-}\rightarrow\rho^{-}\eta_{q}) +\cos\phi \cdot \mathcal{M}(B^{-}\rightarrow\rho^{-}\eta_{s}).
		\end{eqnarray}
		\begin{eqnarray}
			\mathcal{M}(\bar{B^{0}}\rightarrow\rho^{0}\eta)=\cos\phi\cdot \mathcal{M}(\bar{B^{0}}\rightarrow\rho^{0}\eta_{q}) -\sin\phi \cdot \mathcal{M}(\bar{B^{0}}\rightarrow\rho^{0}\eta_{s}),
		\end{eqnarray}
		\begin{eqnarray}
			\mathcal{M}(\bar{B^{0}}\rightarrow\rho^{0}\eta^{\prime})=\sin\phi\cdot \mathcal{M}(\bar{B^{0}}\rightarrow\rho^{0}\eta_{q}) +\cos\phi \cdot \mathcal{M}(\bar{B^{0}}\rightarrow\rho^{0}\eta_{s}).
		\end{eqnarray}
	\end{widetext}
	
The decay width can be calculated by
	\begin{eqnarray}
		\Gamma(B \rightarrow \rho\eta^{(\prime)}) =\frac{G_F^2 m_B^3}{128 \pi} |\mathcal{M}(B\rightarrow \rho\eta^{(\prime)})|^2.
	\end{eqnarray}
Branching ratios and direct $ CP $ violations are defined as
	\begin{eqnarray}
		Br(B\rightarrow \rho\eta^{(\prime)})=\Gamma(B \rightarrow \rho\eta^{(\prime)})/\Gamma_{B},
	\end{eqnarray}
	\begin{eqnarray}
		&&A_{CP}(B^{+}(B^{0})\rightarrow\rho\eta^{(\prime)})\nonumber\\&&=\dfrac{\Gamma(B^{-}(\bar{B^{0}})\rightarrow\rho\eta^{(\prime)})-\Gamma(B^{+}(B^{0})\rightarrow\rho\eta^{(\prime)})}{\Gamma(B^{-}(\bar{B^{0}})\rightarrow\rho\eta^{(\prime)})+\Gamma(B^{+}(B^{0})\rightarrow\rho\eta^{(\prime)})}.
	\end{eqnarray}

	\section{The Contribution of Next-to-Leading-Order Corrections}
In this section, we consider vertex corrections, quark loops, and magnetic penguins, which are the most important next-to-leading-order(NLO) corrections to the decay amplitudes \cite{LMS2005}. The NLO corrections affect the amplitudes by modifying the Wilson coefficients. We define the combinations of Wilson coefficients as
	\begin{eqnarray}
		&&	a_1(\mu) =C_2(\mu) +\frac{C_1(\mu)}{N_c}, \nonumber\\
		&&	a_2(\mu) =C_1(\mu) +\frac{C_2(\mu)}{N_c}, \nonumber\\
		&&	a_i(\mu) =C_i(\mu) +\frac{C_{i\pm 1}(\mu)}{N_c},
	\end{eqnarray}
	with $ i=3-10 $. When $ i $ is odd (even), take the plus (minus) sign.
	
	\subsection{Vertex corrections}
For vertex corrections, we only consider the contributions to the factorizable diagrams, namely Figs. 1(a) and 1(b), which will change the Wilson coefficients to \cite{QCDf1,QCDf2,QCDf3,LMS2005}
	
	\begin{eqnarray}
		&&	a_1(\mu) \rightarrow a_1(\mu)+ \frac{\alpha_s(\mu)}{4 \pi}C_F\frac{C_1(\mu)}{N_c}V_1(M) ,\nonumber\\
		&&	a_2(\mu) \rightarrow a_2(\mu)+ \frac{\alpha_s(\mu)}{4 \pi}C_F\frac{C_2(\mu)}{N_c}V_2(M) ,%\\
		%	&&	a_i(\mu) \rightarrow a_i(\mu)+ \frac{\alpha_s(\mu)}{4 \pi}C_F\frac{C_{i \pm 1}(\mu)}{N_c}V_i(M) ,\nonumber
	\end{eqnarray}
	\begin{equation}
		a_i(\mu) \rightarrow a_i(\mu)+ \frac{\alpha_s(\mu)}{4 \pi}C_F\frac{C_{i \pm 1}(\mu)}{N_c}V_i(M) ,\nonumber
	\end{equation}
with $ i=3-10 $ and $ M $ represents the meson which is emitted. When $ M $ is a pseudoscalar meson, in the naive dimensional regularization (NDR) scheme the function $V_i(M)$ is given by \cite{QCDf1,QCDf2,QCDf3}
	\begin{widetext}
		\begin{equation}
			V_i(M)=\left\{
			\begin{aligned}
				&\int_{0}^{1}dx \phi_{M}^{A}(x)[12\ln(\frac{m_b}{\mu})-18+g(x)],  &\mathrm{for} \; i =1-4,9,10, \\
				&\int_{0}^{1}dx \phi_{M}^{A}(x)[-12\ln(\frac{m_b}{\mu})+6-g(1-x)], &\mathrm{for} \; i =5,7,   \\
				&\int_{0}^{1}dx \phi_P^{M}(x)[-6+h(1-x)], &\mathrm{for} \; i= 6,8
			\end{aligned}
			\right.
		\end{equation}
where $\phi_{M}^{A}(x)$ and $\phi^P_{M}(x)$ are the twist-2 and twist-3 distribution amplitudes of the emitted meson, respectively. When $ M $ is a vector meson, $\phi_{M}^{A}\;(\phi^P_{M})$ is replaced by $\phi_{M}\;(-\phi^s_{M})  $ \cite{HNli-SM2006}. The  hard kernels $ g(x) $ and $ h(x) $ are defined by
		\begin{equation}
			g(x)=3(\dfrac{1-2x}{1-x}lnx-i\pi)+[2Li_{2}(x)-ln^{2}x+\dfrac{2lnx}{1-x}-(3+2i\pi)lnx-(x\leftrightarrow1-x)],
		\end{equation}
		\begin{equation}
			h(x)=2Li_{2}(x)-ln^{2}x-(1+2i\pi)lnx-(x\leftrightarrow1-x).
		\end{equation}
	\end{widetext}

	\subsection{Quark loops}
The effective Hamiltonian of the virtual quark loops for $b\rightarrow d $ transition is \cite{LMS2005}
	\begin{eqnarray}
		\mathcal{H}_{\mathrm{eff}} = &-&\sum_{q=u,c,t}\sum_{q'}\frac{G_F}{\sqrt{2}}V_{qb}V_{qd}^*\frac{\alpha_s(\mu)}{2\pi}C^{(q)}(\mu,l^2)  \nonumber \\
		&\times& (\bar{d}\gamma_{\sigma}(1-\gamma_5)T^a b)(\bar{q}'\gamma^{\sigma}T^a q'),
	\end{eqnarray}
where $ l^{2} $ is the momentum squared of the virtual gluon. When $ q =u $ or $ c $, the function $C^{(q)}(\mu,l^2)$ is
	\begin{equation}\label{cq}
		C^{(q)}(\mu,l^2) =\left[G^{(q)}(\mu,l^2) -\frac{2}{3} \right]C_2(\mu),
	\end{equation}
and when $ q=t $, the function is
	\begin{eqnarray}\label{ct}
		C^{(t)}(\mu,l^2) &=&\left[  G^{(d)}(\mu,l^2) -\frac{2}{3}\right]C_3(\mu) \nonumber \\
		&&+ \sum_{q''=u,d,s,c}G^{(q'')}(\mu,l^2)\left[C_4(\mu) +C_6(\mu) \right]. \nonumber \\
	\end{eqnarray}
	The function $ G^{(q)}(\mu,l^2)$ corresponding to the quark $q$ is
	\begin{equation}
		G^{(q)}(\mu,l^2) = -4 \int_{0}^{1}dx x(1-x) \ln \frac{m_q^2-x(1-x)l^2-i\varepsilon}{\mu^2},
	\end{equation}
where $m_q$ is the mass of quarks for $q=u,d,s,c$.
	
Due to the topological structure of the quark loop being similar to that of the penguin diagram, its contributions can be absorbed into the Wilson coefficients $ a_{4} $ and $ a_{6} $,
	\begin{equation} \label{l-square}
		a_{4,6}(\mu) \rightarrow a_{4,6}(\mu)+\frac{\alpha_s(\mu)}{9\pi}\sum_{q=u,c,t}\frac{V_{qb}V_{	qd}^*}{V_{tb}V_{td}^*}C^{(q)}(\mu,\left<l^2\right>),
	\end{equation}
where $\left<l^2\right>=m_b^2/4$ is the mean-value of the momentum squared of the virtual gluon, which is a reasonable value in the numerical analysis of $ B $ decays.
	
	\subsection{Magnetic penguins}
	
The effective Hamiltonian for the magnetic penguin in the $ b \rightarrow d\textsl{g} $ transition is
	\begin{equation}\label{H-mpg}
		\mathcal{H}_{\mathrm{eff}}= - \frac{G_F}{\sqrt{2}}V_{tb}V_{td}^*C_{8\textsl{g}}O_{8\textsl{g}},
	\end{equation}
where the magnetic penguin operator is
	\begin{equation}
		O_{8\textsl{g}}=\frac{g}{8 \pi^2}m_b \bar{d}_i \sigma_{\mu \nu}(1+\gamma_5)T_{ij}^a G^{a\mu\nu} b_j.
	\end{equation}
	
Because of the similarity in topological structures, we can also absorb the contributions of the magnetic penguin operator into the Wilson coefficients $ a_{4} $ and $ a_{6} $ \cite{LMS2005}, just like the case of the quark loops,
	\begin{equation}
		a_{4,6}(\mu) \rightarrow a_{4,6}(\mu)-\frac{\alpha_s(\mu)}{9\pi} \frac{2m_B}{\sqrt{\left<l^2\right>}}C_{8\textsl{g}}^{\mathrm{eff}}(\mu),
	\end{equation}
where the effective Wilson coefficient $C_{8\textsl{g}}^{\mathrm{eff}}=C_{8\textsl{g}}+C_5$ \cite{Hamiltanion1996}.
	
	\subsection{Spectator hard scattering mechanism with $g^{*}g^{*} \rightarrow \eta(\eta')$}
In this work, we also consider the contributions of the spectator hard scattering mechanism (SHSM), specifically the contributions from the $g^{*}g^{*} \rightarrow \eta(\eta')$ transition process \cite{DKY1998,AKS1998,DuY1998,MutaY2000,YY2001}. Compared with previous studies, the transverse momenta of quarks and gluons are included in our calculation. Figure 2 illustrates the transition process of $g^{*}g^{*} \rightarrow \eta(\eta')$.
	
	\begin{figure}[bth]
		\epsfig{file=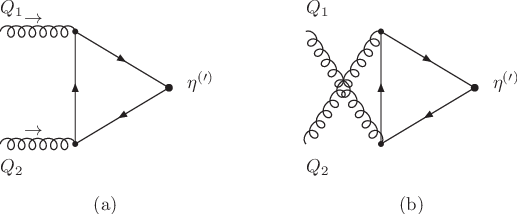,width=7.2cm,height=3.0cm}
		\caption{The Feynman diagrams of $g^*g^*\to\eta(\eta^\prime)$ transition, where diagrams (a) and (b) are two distinct structures.} \label{fig2}
	\end{figure}
	
Mechanism of $g^{*}g^{*} \rightarrow \eta(\eta')$ fusion in $B\to \rho\eta(\eta')$ decays includes two different types of contributions. One is the contribution of diagram of the magnetic penguin operator, and the other is the quark-loop diagram as shown in Fig. 3.
	
	\begin{figure}[bth]
		\epsfig{file=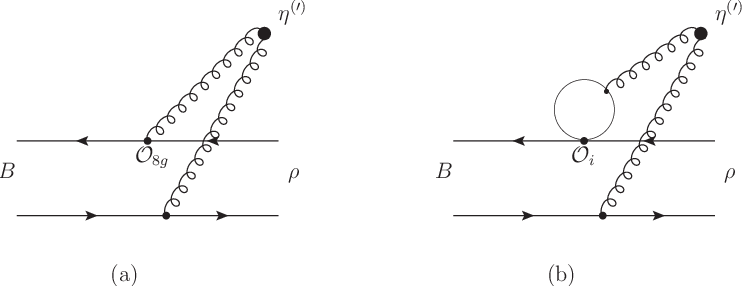,width=8.2cm,height=3.6cm}
		\caption{The Feynman diagrams for $B\rightarrow \rho \eta^{(\prime)}$ decays when considering SHSM, where diagram (a) depicts contribution of the magnetic-penguin operator, and (b) the quark-loop contribution. The solid dots represent $g^{*}g^{*} \rightarrow \eta(\eta')$ transition as shown in Fig. 2.} \label{fig3}
	\end{figure}
	
The amplitude of the magnetic penguin operator is
	\begin{eqnarray}\label{FO8g}
		&F_{O_{8g}}^{q(s)}&=-i\dfrac{2}{N_{c}^{3}}f_{B}f_{\rho}^{||}f_{q(s)}m_{B}^{4}\int_{0}^{1}du\int dx_{1}dx_{2}\int b_{2}db_{2}\nonumber\\&&
		\times b_{3}db_{3}\int_{0}^{2\pi}d\theta\int d\xi dk_{\perp}k_{\perp}(\dfrac{m_{B}}{2}+\dfrac{|\vec{k_{\perp}}|^{2}}{2\xi^{2}m_{B}})K(\vec{k})\nonumber\\&&
		\times (E_{Q}+m_{Q})J_{0}(k_{\perp}b_{3})\alpha_{s}^{2}(\mu^{\prime})(1-x_{1})\phi_{\eta_{q(s)}}(x_{2},b_{2})\nonumber\\&&
		\times \{(E_{q}-k^{3})\phi_{\rho}(x_{1},b_{3})+r_{\rho}r_{f}[x_{1}E_{q}-(x_{1}-2)k^{3}]\nonumber\\&&
		\times \phi_{\rho}^{s}(x_{1},b_{3})+r_{\rho}r_{f}[(x_{1}-2)E_{q}-x_{1}k^{3}]\phi_{\rho}^{t}(x_{1},b_{3}) \}\nonumber\\&&
		\times h^{\prime}(\xi,1-x_{1},1-x_{2},b_{3},b_{2},\theta,u,m_{B})S_{t}(x_{1})\nonumber\\&&
		\times \exp[-S_{B}(\mu^{\prime})-S_{\eta_{q(s)}}(\mu^{\prime})-S_{\rho}(\mu^{\prime})|_{{b_{1}}\rightarrow {b_{3}}}].
	\end{eqnarray}
	
The amplitude of the quark loop is
	\begin{eqnarray}\label{FQL}
		&F_{QL}^{q(s)}&=i\dfrac{2\pi}{N_{c}^{3}}f_{B}f_{\rho}^{||}f_{q(s)}m_{B}^{2}\int_{0}^{1}du\int dx_{1}dx_{2}\int b_{2}db_{2}\nonumber\\&&
		\times b_{3}db_{3}\int_{0}^{2\pi}d\theta\int d\xi dk_{\perp}k_{\perp}(\dfrac{m_{B}}{2}+\dfrac{|\vec{k_{\perp}}|^{2}}{2\xi^{2}m_{B}})K(\vec{k})\nonumber\\&&
		\times(E_{Q}+m_{Q})J_{0}(k_{\perp}b_{3})\alpha_{s}^{2}(\mu^{\prime})(1-x_{1})(1-\xi)\nonumber\\&&
		\times\phi_{\eta_{q(s)}}(x_{2},b_{2})\{(E_{q}-k^{3})\phi_{\rho}(x_{1},b_{3})+2r_{\rho}r_fk^{3}\nonumber\\&&
		\times\phi_{\rho}^{s}(x_{1},b_{3})-2r_{\rho}r_{f}E_{q}\phi_{\rho}^{t}(x_{1},b_{3}) \}\nonumber\\&&
		\times h_{d}(\xi,1-x_{2},1-x_{1},b_{3},b_{2})S_{t}(x_{1})\nonumber\\&&
		\times \exp[-S_{B}(\mu^{\prime})-S_{\eta_{q(s)}}(\mu^{\prime})-S_{\rho}(\mu^{\prime})|_{{b_{1}}\rightarrow {b_{3}}}].
	\end{eqnarray}
	\begin{widetext}
where
		\begin{equation}
			h^{\prime}(\xi,1-x_{1},1-x_{2},b_{3},b_{2},\theta,u,m_{B}) =\left\{
			\begin{aligned}
				\dfrac{\sqrt{b_{2}^{2}+b_{3}^{2}-2b_{2}b_{3}\cos\theta}}{2\sqrt{(1-x_{1})(\xi-u)}m_{B}}&K_{-1}(\sqrt{(1-x_{1})(\xi-u)}m_{B}\sqrt{b_{2}^{2}+b_{3}^{2}-2b_{2}b_{3}\cos\theta})&\\
				\times &K_{0}(-i\sqrt{(1-x_{1})(1-x_{2})}m_{B}b_{2}),&\\
				&  \quad\quad\quad\quad\quad\quad\quad\quad\quad\quad\quad\quad\quad\quad\quad\quad\quad \mathrm{for} \;\xi\geq u, \\
				&&\\
				i\dfrac{\sqrt{b_{2}^{2}+b_{3}^{2}-2b_{2}b_{3}\cos\theta}}{2\sqrt{(1-x_{1})(u-\xi)}m_{B}}&K_{-1}(-i\sqrt{(1-x_{1})(u-\xi)}m_{B}\sqrt{b_{2}^{2}+b_{3}^{2}-2b_{2}b_{3}\cos\theta})&\\
				\times &K_{0}(-i\sqrt{(1-x_{1})(1-x_{2})}m_{B}b_{2}),&\\
				& \quad\quad\quad\quad\quad\quad\quad\quad\quad\quad\quad\quad\quad\quad\quad\quad\quad \mathrm{for} \;\xi\leq u,
			\end{aligned}
			\right.
		\end{equation}
		\begin{equation}
			\mu^{\prime}=\mbox{max}(\sqrt{1-x_{1}}m_{B},\dfrac{1}{b_{3}},\dfrac{1}{b_{2}}).
		\end{equation}
For the decay channels we considered in this work, the additional contributions from the SHSM to decay amplitudes are
		\begin{eqnarray}
			\mathcal{M}_{g^{*}g^{*} \rightarrow \eta(\eta')}(B^{-}\rightarrow\rho^{-}\eta)&=&-V_{t}C^{\mathrm{eff}}_{8g}(\mu^{\prime})(\sqrt{2}\cos\phi\cdot F_{O_{8g}}^{q}-\sin\phi\cdot F_{O_{8g}}^{s})\nonumber\\&&
			+\sum_{q^{\prime}=u,c,t}V_{q^{\prime}}C^{(q^{\prime})}(\mu^{\prime},\langle l^{2}\rangle)(\sqrt{2}\cos\phi\cdot F_{QL}^{q}-\sin\phi\cdot F_{QL}^{s}),
		\end{eqnarray}
		\begin{eqnarray}
			\mathcal{M}_{g^{*}g^{*} \rightarrow \eta(\eta')}(B^{-}\rightarrow\rho^{-}\eta^{\prime})&=&-V_{t}C^{\mathrm{eff}}_{8g}(\mu^{\prime})(\sqrt{2}\sin\phi\cdot F_{O_{8g}}^{q}+\cos\phi\cdot F_{O_{8g}}^{s})\nonumber\\&&
			+\sum_{q^{\prime}=u,c,t}V_{q^{\prime}}C^{(q^{\prime})}(\mu^{\prime},\langle l^{2}\rangle)(\sqrt{2}\sin\phi\cdot F_{QL}^{q}+\cos\phi\cdot F_{QL}^{s}),
		\end{eqnarray}
		\begin{eqnarray}
			\sqrt{2}\mathcal{M}_{g^{*}g^{*} \rightarrow \eta(\eta')}(\bar{B^{0}}\rightarrow\rho^{0}\eta)&=&V_{t}C^{\mathrm{eff}}_{8g}(\mu^{\prime})(\sqrt{2}\cos\phi\cdot F_{O_{8g}}^{q}-\sin\phi\cdot F_{O_{8g}}^{s})\nonumber\\&&
			-\sum_{q^{\prime}=u,c,t}V_{q^{\prime}}C^{(q^{\prime})}(\mu^{\prime},\langle l^{2}\rangle)(\sqrt{2}\cos\phi\cdot F_{QL}^{q}-\sin\phi\cdot F_{QL}^{s}),
		\end{eqnarray}
		\begin{eqnarray}
			\sqrt{2}\mathcal{M}_{g^{*}g^{*} \rightarrow \eta(\eta')}(\bar{B^{0}}\rightarrow\rho^{0}\eta^{\prime})&=&V_{t}C^{\mathrm{eff}}_{8g}(\mu^{\prime})(\sqrt{2}\sin\phi\cdot F_{O_{8g}}^{q}+\cos\phi\cdot F_{O_{8g}}^{s})\nonumber\\&&
			-\sum_{q^{\prime}=u,c,t}V_{q^{\prime}}C^{(q^{\prime})}(\mu^{\prime},\langle l^{2}\rangle)(\sqrt{2}\sin\phi\cdot F_{QL}^{q}+\cos\phi\cdot F_{QL}^{s}),
		\end{eqnarray}
where $ V_{u}=V_{ub}V_{ud}^{*} $, $ V_{c}=V_{cb}V_{cd}^{*} $ and $ V_{t}=V_{tb}V_{td}^{*} $.
	\end{widetext}
	
	\section{The Contribution of Soft Transition Form Factors}
In the numerical calculations, the critical infrared cutoff scale $\mu_c$ needs to be introduced, at which the soft and hard contributions in QCD are separated \cite{LY2021,LY2023,WY2023,LWY2024}. In practice, $\mu_c=1\;\mathrm{GeV}$ is a reasonable value for the infrared cutoff scale \cite{WY2023}. Contributions with energy scale $\mu >\mu_c$ can be calculated with perturbative QCD. While, for contributions with the scale $\mu <\mu_c$, $ B\rho $, $ B\eta_{q} $, $ B\eta_{s} $ soft transition form factors and $ \rho\eta_{q} $, $ \rho\eta_{s} $ soft production form factors are introduced to describe these soft contributions. It has been shown in Ref. \cite{WY2023} that the physical results only slightly depend on the choice of the value of $\mu_c$ around 1 GeV. Therefore, it is reasonable to choose the infrared cutoff scale as $\mu_c=1\;\mathrm{GeV}$.
	
We find that among the eight types of Feynman diagrams shown in Fig. 1, the nonperturbative soft contributions from Figs. 1(c)-(f) are very small and can be neglected. For $B\rightarrow\rho\eta^{(\prime)}$ decays, contributions from the soft production form factors in Figs. 1(g) and 1(h) always cancel between the diagrams interchanging $\rho$ and $\eta^{(\prime)}$ in the final state. Therefore, only the contributions from the soft transition form factors corresponding to Figs. 1(a) and 1(b) remains.
	
Due to the quark composition of $ \eta_{s} $, we only consider the transition form factors of $ B \rightarrow \eta_{q} $ and $ B \rightarrow \rho $. The form factors can be divided into two parts
	\begin{equation} \label{softBM}
		\begin{split}
			&F_+^{B\eta_{q}}=h_+^{B\eta_{q}}+\xi_+^{B\eta_{q}},\\
			&A_0^{B\rho}=h_{A_0}^{B\rho}+\xi_{A_0}^{B\rho},
		\end{split}
	\end{equation}
	% $ A_0^{B\rho} $ is longitudinal transition form factor,
where $h_+^{B\eta_{q}} $ and $ h_{A_0}^{B\rho} $ are the hard form factors that are contributed by hard interactions, and $\xi_+^{B\eta_{q}} $ and $ \xi_{A_0}^{B\rho}$ the soft form factors that are dominated by soft dynamics. Thus, the amplitude is modified as
	\begin{eqnarray}
		&&\mathcal{M}(B^{-}\rightarrow\rho^{-}\eta) \rightarrow \mathcal{M}(B^{-}\rightarrow\rho^{-}\eta)+\cos\phi \dfrac{1}{\sqrt{2}}\nonumber\\&&
		\;\;\;\;\;\times[2if_{\rho}^{||}V_\mathrm{CKM}C_{\eta_{q}\rho}(\mu_{c})\xi_+^{B\eta_{q}}%\nonumber\\&&
		+2if_{q}V_\mathrm{CKM}C_{\rho\eta_{q}}(\mu_{c})\nonumber\\&&
		\;\;\;\;\;\cdot \xi_{A_0}^{B\rho}]-\sin\phi[2if_{s}V_\mathrm{CKM}C_{\rho\eta_{s}}(\mu_{c})%\nonumber\\&&
         \xi_{A_0}^{B\rho}],
	\end{eqnarray}
where $ V_\mathrm{CKM} $ and $ C_{\rho\eta}(\mu_c) $ represent the product of CKM matrix elements and the combination of the Wilson coefficients relevant to the operators $ (V-A)(V-A) $, $ (V-A)(V+A) $ and $ (S+P)(S-P)$, respectively. For $\bar{B^{0}}\rightarrow\rho^{0}\eta  $ decay, $ \cos\phi \dfrac{1}{\sqrt{2}}(\sin\phi) $ should be replaced by $\cos\phi \dfrac{1}{2}(\sin\phi\dfrac{1}{\sqrt{2}})$. When considering decays to $\eta^{\prime}$, only the relevant mixing pattern should be changed.
	
	\section{The Contribution of Color-Octet quark-antiquark pairs in the long-range within the final state}
Since the quark-antiquark pair in mesons should be in color-singlet state, the contributions of quark-antiquark pair in color-octet state in the final state is not considered in $B$ meson decays in general. In principle, as the quark-antiquark pair in color-octet state moves away from each other to the hadronic scale, they can change to color-singlet state by exchanging soft gluons. Therefore, the contributions of color-octet quark-antiquark pairs in the final state of $B$ decays may not be zero. Previously, we have applied the color-octet mechanism to $B\rightarrow PP$ decays \cite{LWY2024}. In this work, we will consider the color-octet contribution in decays of $ B\rightarrow \rho\eta^{(\prime)} $.
	
Utilizing the relation for the generators of the color SU(3) group
	\begin{equation}
		T^{a}_{ij}T^{a}_{kl}=-\dfrac{1}{2N_{c}}\delta_{ij}\delta_{kl}+\dfrac{1}{2}\delta_{il}\delta_{kj},
	\end{equation}
we can separate the contributions of color-singlet and color-octet states \cite{LWY2024}.

\begin{figure}[htb]
		\includegraphics[width=0.40\textwidth]{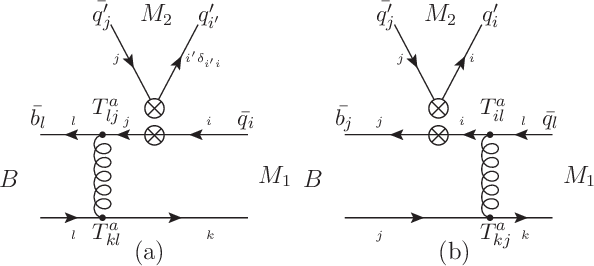}
		\caption{\label{fig-color8} Factorizable diagrams with the insertion of $(\bar{b}_jq_i)(\bar{q}^\prime_i q^\prime_j)$ operator, where $ i,j,k,l $ are color indices, and the specific type of the current in the operators are omitted. (a) is for the diagram with the gluon connecting the antiquark $\bar{b}$ and the light quark line in $B$ meson, and (b) for the gluon connecting the light quark and antiquark line in the meson $M_1$.\label{fig-octet} }
	\end{figure}

Taking Figs. 1(a) and 1(b) as example, we briefly explain the main steps. Let us consider the operator insertion of $(\bar{b}_jq_i)(\bar{q}^\prime_i q^\prime_j)$ in Fig. \ref{fig-octet}.
	
The color factors in Fig. 4(a) is
	\begin{equation}\label{eq:cfa}
		\begin{split}
			\sum_{ijkli^{\prime}}T_{kl}^a T_{lj}^a \delta_{i^\prime i}
			&= \sum_{ijki^\prime} C_F \delta_{kj} \delta_{i^\prime i} \\
			& = \sum_{ijki^\prime} C_F \left(\dfrac{1}{N_c} \delta_{ki} \delta_{i^\prime j} + 2 T_{ki}^a T_{i^\prime j}^a\right),
		\end{split}
	\end{equation}
while for Fig. 4(b), the color factor becomes
	\begin{equation}\label{eq:cfb}
		\begin{split}
			&\sum_{ijkl} T_{il}^a T_{kj}^a
			= \sum_{ijkl} \left[-\dfrac{1}{2N_c} \delta_{il} \delta_{kj} + \dfrac{1}{2} \delta_{ij} \delta_{kl} \right] \\
			&\quad = \sum_{ijkl} \left[-\dfrac{1}{2N_c} \left(\dfrac{1}{N_c} \delta_{ij} \delta_{kl}
			+ 2 T_{ij}^b T_{kl}^b\right) + \dfrac{1}{2} \delta_{ij} \delta_{kl} \right]\\
			&\quad = \sum_{ijkl} \left(\dfrac{C_F}{N_c} \delta_{ij} \delta_{kl}
			- \dfrac{1}{N_c} T_{ij}^b T_{kl}^b \right),
		\end{split}
	\end{equation}
where the first and second terms represent the contributions of color-singlet and color-octet states, respectively. If we consider the insertion of the operator $(\bar{b}_iq_i)(\bar{q}^\prime_j q^\prime_j)$ in Fig. \ref{fig-octet}, there are no contributions of the color-octet state. In addition, we need to introduce phenomenological parameters $ Y^{8}_{F} $ and $ Y^{8}_{M} $ to characterize the magnitude of  color-octet contributions for factorizable and nonfactorizable diagrams, respectively. For the Figs. \ref{fig-octet}(a) and \ref{fig-octet}(b), the result is
	\begin{equation}\label{nfFe8}
		Y^8_F F_e^{(R,P)8},
	\end{equation}
where $ F_e^{(R,P)8} \equiv 2N_c^2 F_e^{(R,P)a} - \dfrac{N_c}{C_F} F_e^{(R,P)b} $. The symbol without the superscript $(R,P)$ represents the result of $ (V-A)(V-A) $ current, while the superscript $ R $ and $ P $ represent the currents of $ (V-A)(V+A) $ and $ (S+P)(S-P) $ operators, respectively. Here the distribution amplitudes of quark-antiquark pairs in color-octet state are assumed as the same as that of color-singlet state.
	
In $ B\rightarrow\rho\eta^{(\prime)} $ decays, there are no extra color-octet contributions from Figs. 1(g) and 1(h), all the soft contributions can be included into the soft form factors. The contributions of Figs. 1(c), 1(d), 1(e) and 1(f) are
	\begin{equation}
		Y^8_M M_e^{(R,P)\prime8},\;Y^8_M M_e^{(R,P) 8},\;Y^8_M M_a^{(R,P)8},
	\end{equation}
	where
	\begin{eqnarray}
		&&M_e^{(R,P)\prime 8} \equiv \dfrac{N_c^2}{C_F} M_e^{(R,P)},
		M_a^{(R,P)8} \equiv -\dfrac{N_c}{C_F} M_a^{(R,P)},\nonumber\\&&
		M_e^{(R,P)8} \equiv 2N_c^2 M_e^{(R,P)c} - \dfrac{N_c}{C_F} M_e^{(R,P)d},%\nonumber\\&&
	\end{eqnarray}
where $ F_e^{(R,P)a} $, $ F_e^{(R,P)b} $, $ M_e^{(R,P)} $, $ M_e^{(R,P)c} $, $ M_e^{(R,P)d} $,  and $ M_a^{(R,P)} $ are the hard amplitudes given in Eqs. (\ref{Fe})$ - $(\ref{MaP-eta}). The decay amplitudes in Eqs. (\ref{Mrho-etaq})$ - $(\ref{Mrho0baretas}) are modified as
	\begin{widetext}
		\begin{eqnarray}
			\sqrt{2}\mathcal{M}(B^{-}\rightarrow\rho^{-}\eta_{q}) &\rightarrow& \sqrt{2}\mathcal{M}(B^{-}\rightarrow\rho^{-}\eta_{q})
			+[V_{u}{C_{2}}-V_{t}(C_{3}+2C_{4}-\dfrac{1}{2}C_{9}+\dfrac{1}{2}C_{10})]Y^{8\rho\eta_{q}}_{F}F^{8}_{e\rho\eta_{q}}\nonumber\\&& -V_{t}(2C_{6}+\dfrac{1}{2}C_{8})Y^{8\rho\eta_{q}}_{F}F_{e\rho\eta_{q}}^{R8}-V_{t}(C_{5}-\dfrac{1}{2}C_{7})Y^{8\rho\eta_{q}}_{F}F_{e\rho\eta_{q}}^{P8}\nonumber\\&&
			+[V_{u}C_{1}-V_{t}(C_{3}+C_{9})]Y^{8\rho\eta_{q}}_{F}F^{8}_{e\eta_{q}\rho}-V_{t}(C_{5}+C_{7})Y^{8\rho\eta_{q}}_{F}F_{e\eta_{q}\rho}^{P8}\nonumber\\&&
			+[V_{u}C_{2}-V_{t}(C_{3}+2C_{4}-\dfrac{1}{2}C_{9}+\dfrac{1}{2}C_{10})]Y^{8\rho\eta_{q}}_{M}M^{8}_{e\rho\eta_{q}}-V_{t}(C_{5}-\dfrac{1}{2}C_{7})Y^{8\rho\eta_{q}}_{M}M_{e\rho\eta_{q}}^{R8}\nonumber\\&&
			+[V_{u}C_{1}-V_{t}(C_{4}+2C_{3}-\dfrac{1}{2}C_{10}+\dfrac{1}{2}C_{9})]Y^{8\rho\eta_{q}}_{M}M^{\prime8}_{e\rho\eta_{q}}-V_{t}(C_{6}-\dfrac{1}{2}C_{8})Y^{8\rho\eta_{q}}_{M}M_{e\rho\eta_{q}}^{R\prime8}\nonumber\\&& -V_{t}(2C_{6}+\dfrac{1}{2}C_{8})Y^{8\rho\eta_{q}}_{M}M_{e\rho\eta_{q}}^{P8}-V_{t}(2C_{5}+\dfrac{1}{2}C_{7})Y^{8\rho\eta_{q}}_{M}M_{e\rho\eta_{q}}^{P\prime8}\nonumber\\&&
			+[V_{u}C_{1}-V_{t}(C_{3}+C_{9})]Y^{8\rho\eta_{q}}_{M}M^{8}_{e\eta_{q}\rho}+[V_{u}C_{2}-V_{t}(C_{4}+C_{10})]Y^{8\rho\eta_{q}}_{M}M^{\prime8}_{e\eta_{q}\rho}\nonumber\\&&
			-V_{t}(C_{5}+C_{7})Y^{8\rho\eta_{q}}_{M}M_{e\eta_{q}\rho}^{R8}-V_{t}(C_{6}+C_{8})Y^{8\rho\eta_{q}}_{M}M_{e\eta_{q}\rho}^{R\prime8}\nonumber\\&&
			+[V_{u}C_{1}-V_{t}(C_{3}+C_{9})]Y^{8\rho\eta_{q}}_{M}(M^{8}_{a\rho\eta_{q}}+M^{8}_{a\eta_{q}\rho})-V_{t}(C_{5}+C_{7})Y^{8\rho\eta_{q}}_{M}(M_{a\rho\eta_{q}}^{R8}+M_{a\eta_{q}\rho}^{R8}),
		\end{eqnarray}
		\begin{eqnarray}
			\mathcal{M}(B^{-}\rightarrow\rho^{-}\eta_{s}) &\rightarrow& \mathcal{M}(B^{-}\rightarrow\rho^{-}\eta_{q})
			-V_{t}(C_{4}-\dfrac{1}{2}C_{10})Y^{8\rho\eta_{s}}_{F}F^{8}_{e\rho\eta_{s}}-V_{t}(C_{6}-\dfrac{1}{2}C_{8})Y^{8\rho\eta_{s}}_{F}F_{e\rho\eta_{s}}^{R8}\nonumber\\&&-V_{t}(C_{4}-\dfrac{1}{2}C_{10})Y^{8\rho\eta_{s}}_{M}M^{8}_{e\rho\eta_{s}}-V_{t}(C_{3}-\dfrac{1}{2}C_{9})Y^{8\rho\eta_{s}}_{M}M^{\prime8}_{e\rho\eta_{s}}\nonumber\\&&
			-V_{t}(C_{6}-\dfrac{1}{2}C_{8})Y^{8\rho\eta_{s}}_{M}M_{e\rho\eta_{s}}^{P8}-V_{t}(C_{5}-\dfrac{1}{2}C_{7})Y^{8\rho\eta_{s}}_{M}M_{e\rho\eta_{s}}^{P\prime8},
		\end{eqnarray}
		\begin{eqnarray}
			2\mathcal{M}(\bar{B^{0}}\rightarrow\rho^{0}\eta_{q}) &\rightarrow& 2\mathcal{M}(\bar{B^{0}}\rightarrow\rho^{0}\eta_{q})
			-[V_{u}{C_{2}}-V_{t}(C_{3}+2C_{4}-\dfrac{1}{2}C_{9}+\dfrac{1}{2}C_{10})]Y^{8\rho\eta_{q}}_{F}F^{8}_{e\rho\eta_{q}}\nonumber\\&&
			+V_{t}(2C_{6}+\dfrac{1}{2}C_{8})Y^{8\rho\eta_{q}}_{F}F_{e\rho\eta_{q}}^{R8}+V_{t}(C_{5}-\dfrac{1}{2}C_{7})Y^{8\rho\eta_{q}}_{F}F_{e\rho\eta_{q}}^{P8}-V_{t}\dfrac{3}{2}C_{8}Y^{8\rho\eta_{q}}_{F}F_{e\eta_{q}\rho}^{R8}\nonumber\\&&
			+[V_{u}C_{2}-V_{t}(-C_{3}+\dfrac{1}{2}C_{9}+\dfrac{3}{2}C_{10})]Y^{8\rho\eta_{q}}_{F}F^{8}_{e\eta_{q}\rho}+V_{t}(C_{5}-\dfrac{1}{2}C_{7})Y^{8\rho\eta_{q}}_{F}F_{e\eta_{q}\rho}^{P8}\nonumber\\&&
			-[V_{u}C_{2}-V_{t}(C_{3}+2C_{4}-\dfrac{1}{2}C_{9}+\dfrac{1}{2}C_{10})]Y^{8\rho\eta_{q}}_{M}M^{8}_{e\rho\eta_{q}}+V_{t}(C_{5}-\dfrac{1}{2}C_{7})Y^{8\rho\eta_{q}}_{M}M_{e\rho\eta_{q}}^{R8}\nonumber\\&&
			-[V_{u}C_{1}-V_{t}(C_{4}+2C_{3}-\dfrac{1}{2}C_{10}+\dfrac{1}{2}C_{9})]Y^{8\rho\eta_{q}}_{M}M^{\prime8}_{e\rho\eta_{q}}+V_{t}(C_{6}-\dfrac{1}{2}C_{8})Y^{8\rho\eta_{q}}_{M}M_{e\rho\eta_{q}}^{R\prime8}\nonumber\nonumber\\&&
			+V_{t}(2C_{6}+\dfrac{1}{2}C_{8})Y^{8\rho\eta_{q}}_{M}M_{e\rho\eta_{q}}^{P8}+V_{t}(2C_{5}+\dfrac{1}{2}C_{7})Y^{8\rho\eta_{q}}_{M}M_{e\rho\eta_{q}}^{P\prime8}\nonumber\\&&
			+[V_{u}C_{2}-V_{t}(-C_{3}+\dfrac{1}{2}C_{9}+\dfrac{3}{2}C_{10})]Y^{8\rho\eta_{q}}_{M}M^{8}_{e\eta_{q}\rho}\nonumber+V_{t}(C_{5}-\dfrac{1}{2}C_{7})Y^{8\rho\eta_{q}}_{M}M_{e\eta_{q}\rho}^{R8}\\&&
			+[V_{u}C_{1}-V_{t}(-C_{4}+\dfrac{1}{2}C_{10}+\dfrac{3}{2}C_{9})]Y^{8\rho\eta_{q}}_{M}M^{\prime8}_{e\eta_{q}\rho}+V_{t}(C_{6}-\dfrac{1}{2}C_{8})Y^{8\rho\eta_{q}}_{M}M_{e\eta_{q}\rho}^{R\prime8}\nonumber\\&&
			-V_{t}\dfrac{3}{2}C_{8}Y^{8\rho\eta_{q}}_{M}M_{e\eta_{q}\rho}^{P8}-V_{t}\dfrac{3}{2}C_{7}Y^{8\rho\eta_{q}}_{M}M_{e\eta_{q}\rho}^{P\prime8}\nonumber\\&&
			+[V_{u}C_{2}-V_{t}(-C_{3}+\dfrac{1}{2}C_{9}+\dfrac{3}{2}C_{10})]Y^{8\rho\eta_{q}}_{M}(M^{8}_{a\rho\eta_{q}}+M^{8}_{a\eta_{q}\rho})\nonumber\\&&
			+V_{t}(C_{5}-\dfrac{1}{2}C_{7})Y^{8\rho\eta_{q}}_{M}(M_{a\rho\eta_{q}}^{R8}+M_{a\eta_{q}\rho}^{R8})-V_{t}\dfrac{3}{2}C_{8}Y^{8\rho\eta_{q}}_{M}(M_{a\rho\eta_{q}}^{P8}+M_{a\eta_{q}\rho}^{P8}),
		\end{eqnarray}
		\begin{eqnarray}
			\sqrt{2}\mathcal{M}(\bar{B^{0}}\rightarrow\rho^{0}\eta_{s}) &\rightarrow& \sqrt{2}\mathcal{M}(\bar{B^{0}}\rightarrow\rho^{0}\eta_{s})
			+V_{t}(C_{4}-\dfrac{1}{2}C_{10})Y^{8\rho\eta_{s}}_{F}F^{8}_{e\rho\eta_{s}}+V_{t}(C_{6}-\dfrac{1}{2}C_{8})Y^{8\rho\eta_{s}}_{F}F_{e\rho\eta_{s}}^{R8}\nonumber\\&&
			+V_{t}(C_{4}-\dfrac{1}{2}C_{10})Y^{8\rho\eta_{s}}_{M}M^{8}_{e\rho\eta_{s}}+V_{t}(C_{3}-\dfrac{1}{2}C_{9})Y^{8\rho\eta_{s}}_{M}M^{\prime8}_{e\rho\eta_{s}}\nonumber\\&&
			+V_{t}(C_{6}-\dfrac{1}{2}C_{8})Y^{8\rho\eta_{s}}_{M}M_{e\rho\eta_{s}}^{P8}+V_{t}(C_{5}-\dfrac{1}{2}C_{7})Y^{8\rho\eta_{s}}_{M}M_{e\rho\eta_{s}}^{P\prime8}.
		\end{eqnarray}
	\end{widetext}

	\section{Numerical Result and Discussion}
In the numerical calculations, besides the parameters in the meson wave functions, the other nonperturbative parameters are the soft transition form factors $\xi_{+}^{B\eta_{q}}$, $ \xi_{A_0}^{B\rho} $ and the color-octet parameters $Y^{8\rho\eta_{q}}_{F}$, $Y^{8\rho\eta_{s}}_{F}$, $Y^{8\rho\eta_{q}}_{M}$, and $Y^{8\rho\eta_{s}}_{M}$.
	
At the energy scale $ \mu>\mu_{c} $, the hard transition form factors can be calculated by using PQCD approach, we have
	\begin{eqnarray}
		&&h_{+}^{B\eta_q} = 0.17 \pm 0.01,  \nonumber\\
		&&h_{A_0}^{B\rho}  \ = 0.18 \pm 0.01.
	\end{eqnarray}
From the experimental data of semileptonic decays of $ B $ meson, as well as the calculation results from nonperturbative method, such as light-cone sum rules (LCSR), etc., the complete transition form factors can be extracted \cite{PDG2024,PB-RZ2005,AK-TM-NO2007,AB-DMS-RZ2021,NG-AK-DVD2020,JG-CDL-YLS-YMW-YBW2020,MAI-JGK-SGK-CDR2007,MAI-JGK-SGK-PS-GGS2012}
	\begin{eqnarray}\label{formA}
		&&F_{+}^{B\eta_q} = 0.23 \pm 0.03, \nonumber\\
		&&A_0^{B\rho} \ = 0.32 \pm 0.05.
	\end{eqnarray}
The value of $ F_+^{B\eta_q} $ is obtained by using the experimental data on the branching ratios of semileptonic decays of $ B $ meson \cite{PDG2024}. For the numerical value of $ A_0^{B\rho} $, there are several theoretical results from LCSR, soft-collinear effective theory (SCET) and quark model \cite{PB-RZ2005,AK-TM-NO2007,AB-DMS-RZ2021,NG-AK-DVD2020,JG-CDL-YLS-YMW-YBW2020,MAI-JGK-SGK-CDR2007,MAI-JGK-SGK-PS-GGS2012}. Here the value of  $ A_0^{B\rho} $ given in Eq. (\ref{formA}) is obtained by averaging the results from LCSR method \cite{PB-RZ2005,AK-TM-NO2007,AB-DMS-RZ2021,NG-AK-DVD2020}, which is very close to that from SCET and quark model given in Refs. \cite{JG-CDL-YLS-YMW-YBW2020,MAI-JGK-SGK-CDR2007,MAI-JGK-SGK-PS-GGS2012}.
	
Based on the above results, we can obtain the values of soft transition form factors
	\begin{eqnarray}
		&&\xi_{+}^{B\eta_q} = 0.06 \pm0.02, \nonumber\\
		&&\xi_{A_0}^{B\rho} \ = 0.14 \pm  0.04 .
	\end{eqnarray}
	
For the color-octet parameters, since perturbative method cannot be applied for calculating them, their values are determined by fitting the experimental data of $ B\rightarrow \rho\eta^{(\prime)}$ decays. We find that there are two different parameter solutions that can all yield the best results, and we denote them as $ a $ and $ b $. The obtained results are given in Table I.
	
	\begin{table*}[htb]
		\renewcommand\arraystretch{1.5}
		\caption{The values of color-octet parameters.}
		\begin{threeparttable}
			\setlength{\tabcolsep}{3.05mm}{
				\begin{tabular}{c c c c c}
					\hline
					\hline
					& $ Y^{8\rho\eta_{q}}_{F} $ & $ Y^{8\rho\eta_{s}}_{F} $ & $ Y^{8\rho\eta_{q}}_{M} $ &  $ Y^{8\rho\eta_{s}}_{M} $  \\
					\hline
					$ a $ & $ (0.171^{+0.039}_{-0.078})e^{i\pi(1.394^{+0.106}_{-0.058})} $ & $ (0.172^{+0.038}_{-0.075})e^{i\pi(1.625^{+0.159}_{-0.284})} $  & $ (0.187^{+0.020}_{-0.038})e^{i\pi(1.159^{+0.054}_{-0.043})} $ & $ (0.196^{+0.013}_{-0.031})e^{i\pi(0.860^{+0.114}_{-0.057})} $  \\
					
					$ b $ & $ (0.163^{+0.009}_{-0.008})e^{i\pi(0.174^{+0.032}_{-0.028})} $ & $ (0.017^{+0.019}_{-0.014})e^{i\pi(0.289^{+0.146}_{-0.204})} $  & $ (0.165^{+0.008}_{-0.004})e^{i\pi(0.203^{+0.029}_{-0.030})} $ & $ (0.205^{+0.002}_{-0.002})e^{i\pi(0.286^{+0.021}_{-0.019})} $  \\
					
					\hline
					\hline
			\end{tabular}  }
		\end{threeparttable}
	\end{table*}
	
The comparison of the theoretical results and experimental data for the branching ratios and direct $ CP $ violations of $ B\rightarrow \rho\eta^{(\prime)} $ decays are presented in Table II.
	
Column ``$\mathrm{LO_{NLOWC}}$" in Table II represents the leading-order contribution of QCD with NLO Wilson coefficients. Column ``NLO" represents the contribution up to next-to-leading-order in QCD including the vertex corrections, quark loops and magnetic penguin contributions. In column ``NLO+$ g^{*}g^{*} $", the contribution of SHSM is included. In column ``NLO+$ g^{*}g^{*}+\xi^{B\rho}+\xi^{B\eta_{q}} $", the contributions of NLO,  SHSM of $ g^{*}g^{*}$ fusion and soft transition form factors are included. ``soft" denotes the contributions of soft transition form factors and color-octet parameters.
	
The first uncertainty in the theoretical result comes from the uncertainties of soft transition form factors and color-octet parameters. The second and third uncertainties originate from the uncertainties of the parameters in $B$ and light meson wave functions, respectively.
	
	\begin{table*}[htb]
		\renewcommand\arraystretch{1.5}
		\caption{$ B\rightarrow \rho\eta^{(\prime)} $ branching ratios and $ CP $ violations.
			\label{table-BRCP}}
		\begin{threeparttable}
			\setlength{\tabcolsep}{0.15mm}{
				\begin{tabular}{c c c c c c c c}
					\hline
					\hline
					&$\mathrm{LO_{NLOWC}}$	 & NLO	 & NLO+$ g^{*}g^{*} $  & $ \begin{aligned}&\text{NLO}+ g^{*}g^{*}\\&+ \xi^{B\rho}+\xi^{B\eta_{q}} \end{aligned}$ & NLO+$ g^{*}g^{*} $+soft$ ^{a} $ & NLO+$ g^{*}g^{*} $+soft$ ^{b} $ &  Data \cite{PDG2024}    \\
					\hline
					Br($ B^{0}\rightarrow \rho^{0}\eta $)$ \times 10^{-6} $         &0.002  &0.01  &0.01  &0.05  &$ 0.15\pm0.09^{+0.00+0.00}_{-0.00-0.00} $&$ 0.28\pm0.04^{+0.00+0.00}_{-0.00-0.00} $& $ < $1.5        \\
					Br($ B^{+}\rightarrow \rho^{+}\eta $)$ \times 10^{-6} $         &3.45   &3.66  &3.68  &6.49  &$ 9.35\pm2.02^{+0.14+0.29}_{-0.23-0.29} $&$ 9.42\pm1.51^{+0.14+0.29}_{-0.23-0.29} $& 7.0$ \pm $2.9   \\
					Br($ B^{0}\rightarrow \rho^{0}\eta^{\prime} $)$ \times 10^{-6} $&0.01   &0.01  &0.02  &0.07  &$ 1.21\pm0.20^{+0.00+0.00}_{-0.00-0.00} $&$ 1.26\pm0.08^{+0.00+0.00}_{-0.00-0.00} $& $ < $1.3        \\
					Br($ B^{+}\rightarrow \rho^{+}\eta^{\prime} $)$ \times 10^{-6} $&2.08   &1.84  &1.94  &2.58  &$ 7.83\pm1.38^{+0.07+0.14}_{-0.11-0.14} $&$ 7.69\pm1.07^{+0.07+0.14}_{-0.11-0.14} $& 9.7$ \pm $2.2   \\
					\hline
					\hline
					$ A_{CP} $($ B^{0}\rightarrow \rho^{0}\eta $)                   &0.94  &0.84     &0.82     &0.22    &$-0.72\pm0.35^{+0.01+0.07}_{-0.01-0.09} $ &$-0.51\pm0.30^{+0.01+0.07}_{-0.01-0.09} $&  - \\
					$ A_{CP} $($ B^{+}\rightarrow \rho^{+}\eta $)                   &0.00  &$-$0.08  &$-$0.09  &$-$0.14 &$ 0.14\pm0.08^{+0.00+0.00}_{-0.00-0.00} $ &$ 0.15\pm0.07^{+0.00+0.00}_{-0.00-0.00} $   &  0.11$ \pm $0.11 \\
					$ A_{CP} $($ B^{0}\rightarrow \rho^{0}\eta^{\prime} $)          &0.69  &0.67     &0.62     &0.04    &$-0.29\pm0.33^{+0.02+0.08}_{-0.03-0.09} $ &$-0.35\pm0.33^{+0.02+0.08}_{-0.03-0.09} $&  - \\
					$ A_{CP} $($ B^{+}\rightarrow \rho^{+}\eta^{\prime} $)          &0.10  &0.20     &0.16     &0.34    &$ 0.14\pm0.11^{+0.00+0.03}_{-0.00-0.03} $ &$ 0.17\pm0.11^{+0.00+0.03}_{-0.00-0.03} $   &  0.26$ \pm $0.17 \\
					\hline
					\hline
			\end{tabular}  }
		\end{threeparttable}	
	\end{table*}
	
Comparing the results in the first and second columns in Table \ref{table-BRCP}, one can see that the NLO contributions only slightly affect the decay amplitudes. The branching ratios are only changed by less than 10\% for most decay modes. Only for $ B^{0}\rightarrow \rho^{0}\eta $ decay, the effect of NLO contribution is dramatically large. This is because the leading order contribution is doubly suppressed by the color and isospin structures of the tree-level diagrams for this decay mode. The SHSM effect from $g^*g^*$ fusion process is also very small. Only the soft transition form factors and especially the color-octet contributions can enhance the branching ratios effectively, which is important for explaining the experimental data. For the experimental measured decay modes, the calculated branching ratios and $CP$ violations can be in good agreement with experimental data. We also predict the branching ratios and $CP$ violations for $B^{0}\rightarrow \rho^{0}\eta$ and $\rho^{0}\eta^{\prime}$ decays, which have not been well measured yet in experiment. These predictions can be tested in experiment in the future.
	
	\section{Summary}
We study $B\rightarrow \rho\eta$, $\rho\eta^{\prime}$ decays in the modified perturbative QCD approach, where a critical infrared cutoff scale $\mu_c$ is introduced. For contributions at scales larger than  $\mu_c$, PQCD approach can be applied. For contributions at scale $\mu<\mu_c$, soft transition form factors are introduced to describe these contributions. In addition, color-octet contribution is introduced, where the quark-antiquark pairs in $B$ decay can be in color-octet state after short-distance interactions, then color-octet quark-antiquark pairs can be changed to color-singlet state by exchanging soft gluons at long-distance of hadronic scale. By selecting reasonable input parameters, we find the experimental data can be well explained by the modified PQCD approach. We also predict branching ratios and $CP$ violations for two decay modes which have not been well detected in experiment. These predictions can be tested in experiment in future.

	\vspace{0.5cm}
	\acknowledgments
This work is supported in part by the National Natural Science Foundation of China under Contracts No. 12275139, 11875168.

    \begin{center}
    	\textbf{DATA AVAILABILITY}
    \end{center}
    
The data that support the findings of this article are openly available \cite{PDG2024}.
	
	\appendix{}
	\section{\label{a}THRESHOLD AND SUDAKOV FACTOR}	
Threshold factor $S_t(x)$ can be found in Ref. \cite{lihn2002} as
	\begin{equation}
		S_t(x)=\frac{2^{1+2c}\Gamma(3/2+c)}{\sqrt{\pi}\Gamma(1+c)}[x(1-x)]^c,
	\end{equation}
where $c=0.3$.
	
The exponentials $\exp[-S_{B,\eta_{q(s)},\rho}(\mu)]$ include the Sudakov factors associated with the mesons and the relevant single ultraviolet logarithms. The exponent parts are
	\begin{equation}
		S_B(\mu) = s(x,b,m_B)-\frac{1}{\beta_1}\ln \frac{\ln (\mu/\Lambda_{\mbox{QCD}})}
		{\ln (1/(b\Lambda_{\mbox{QCD}}))},
	\end{equation}
	\begin{eqnarray}
		S_{\eta_{q(s)},\rho}(\mu) &=& s(x,b,m_B)+s(1-x,b,m_B)\nonumber\\
		&&\;\;-\frac{1}{\beta_1}\ln \frac{\ln (\mu/\Lambda_{\mbox{QCD}})}
		{\ln (1/(b\Lambda_{\mbox{QCD}}))},
	\end{eqnarray}
where $s(x,b,Q)$ up to next-to-leading order is \cite{Li1995}
	\begin{widetext}
		\begin{eqnarray}
			&& s(x,b,Q)=\frac{A^{(1)}}{2\beta_{1}}\hat{q}\ln\left(\frac{\hat{q}}
			{\hat{b}}\right)-
			\frac{A^{(1)}}{2\beta_{1}}\left(\hat{q}-\hat{b}\right)+
			\frac{A^{(2)}}{4\beta_{1}^{2}}\left(\frac{\hat{q}}{\hat{b}}-1\right)
			-\left[\frac{A^{(2)}}{4\beta_{1}^{2}}-\frac{A^{(1)}}{4\beta_{1}}
			\ln\left(\frac{e^{2\gamma_E-1}}{2}\right)\right]
			\ln\left(\frac{\hat{q}}{\hat{b}}\right)
			\nonumber \\
			&&+\frac{A^{(1)}\beta_{2}}{4\beta_{1}^{3}}\hat{q}\left[
			\frac{\ln(2\hat{q})+1}{\hat{q}}-\frac{\ln(2\hat{b})+1}{\hat{b}}\right]
			+\frac{A^{(1)}\beta_{2}}{8\beta_{1}^{3}}\left[
			\ln^{2}(2\hat{q})-\ln^{2}(2\hat{b})\right]
			\nonumber \\
			&&+\frac{A^{(1)}\beta_{2}}{8\beta_{1}^{3}}
			\ln\left(\frac{e^{2\gamma_E-1}}{2}\right)\left[
			\frac{\ln(2\hat{q})+1}{\hat{q}}-\frac{\ln(2\hat{b})+1}{\hat{b}}\right]
			-\frac{A^{(1)}\beta_{2}}{16\beta_{1}^{4}}\left[
			\frac{2\ln(2\hat{q})+3}{\hat{q}}-\frac{2\ln(2\hat{b})+3}{\hat{b}}\right]
			\nonumber \\
			& &-\frac{A^{(1)}\beta_{2}}{16\beta_{1}^{4}}
			\frac{\hat{q}-\hat{b}}{\hat{b}^2}\left[2\ln(2\hat{b})+1\right]
			+\frac{A^{(2)}\beta_{2}^2}{432\beta_{1}^{6}}
			\frac{\hat{q}-\hat{b}}{\hat{b}^3}
			\left[9\ln^2(2\hat{b})+6\ln(2\hat{b})+2\right]
			\nonumber \\
			&& +\frac{A^{(2)}\beta_{2}^2}{1728\beta_{1}^{6}}\left[
			\frac{18\ln^2(2\hat{q})+30\ln(2\hat{q})+19}{\hat{q}^2}
			-\frac{18\ln^2(2\hat{b})+30\ln(2\hat{b})+19}{\hat{b}^2}\right]
			\label{sss},
		\end{eqnarray}
with $\hat q$ and $\hat b$ being defined as
		\begin{equation}
			{\hat q} \equiv  {\rm ln}\left(xQ/(\sqrt 2\Lambda_{QCD})\right),~
			{\hat b} \equiv  {\rm ln}(1/b\Lambda_{QCD}).
		\end{equation}
The coefficients $\beta_{i}$ and $A^{(i)}$ are
		\begin{eqnarray}
			& &\beta_{1}=\frac{33-2n_{f}}{12}\;,\;\;\;\beta_{2}=\frac{153-19n_{f}}{24}\; ,
			A^{(1)}=\frac{4}{3}\;,
			\nonumber \\
			& & A^{(2)}=\frac{67}{9}-\frac{\pi^{2}}{3}-\frac{10}{27}n_
			{f}+\frac{8}{3}\beta_{1}\ln\left(\frac{e^{\gamma_E}}{2}\right)\;,
		\end{eqnarray}
where $\gamma_E$ is the Euler constant.
	\end{widetext}
	
	\section{\label{b}LIGHT MESON DISTRIBUTION AMPLITUDES}
The light-cone distribution amplitudes of the light mesons are $\phi_{\eta_{q(s)}}(x,k_{\perp})$, $\phi^{\eta_{q(s)}}_P(x,k_{\perp})$, $\phi^{\eta_{q(s)}}_\sigma(x,k_{\perp})$, $\phi_{\rho}(x,k_{\perp})$, $\phi_{\rho}^{t}(x,k_{\perp})$ and $\phi_{\rho}^{s}(x,k_{\perp})$. Assuming the dependence of transverse momentum takes the form of Gaussian distribution, when transforming the function to b-space, we have \cite{wy2002}
	\begin{eqnarray}
		\phi(x,b)=\phi(x)\exp\left(-\frac{b^2}{4\beta^2}\right).
	\end{eqnarray}
For the distribution amplitudes of $ \eta_{q(s)} $ and $ \rho $ mesons we use $\beta=4.0 \;\mathrm{GeV}^{-1} $ \cite{wy2002,JK93}.
	\begin{widetext}
		For $ \eta_{q(s)} $ meson, the twist-2 and twist-3 distribution amplitudes are given by \cite{Ball-Braun2006}
		\begin{equation}
			\phi_{\eta_{q(s)}}(x)=6x(1-x)\biggl[1+a_1^{\eta_{q(s)}} C_1^{3/2}(t)+a_2^{\eta_{q(s)}} C_2^{3/2}(t)\biggr],
		\end{equation}
		\begin{equation}
			\phi_P^{\eta_{q(s)}}(x)=1+a_{0P}^{\eta_{q(s)}}+a_{1P}^{\eta_{q(s)}}C_1^{1/2}(t)+a_{2P}^{\eta_{q(s)}}C_2^{1/2}(t)+a_{3P}^{\eta_{q(s)}}C_3^{1/2}(t) +a_{4P}^{\eta_{q(s)}}C_4^{1/2}(t) +b_{1P}^{\eta_{q(s)}}\ln(x)+b_{2P}^{\eta_{q(s)}}\ln(1-x),
		\end{equation}
		\begin{equation}
			\begin{split}
				\phi_\sigma^{\eta_{q(s)}}(x)&=6x(1-x)\biggl[1+a_{0\sigma}^{\eta_{q(s)}}+a_{1\sigma}^{\eta_{q(s)}}C_1^{3/2}(t) +a_{2\sigma}^{\eta_{q(s)}}C_2^{3/2}(t)+a_{3\sigma}^{\eta_{q(s)}}C_3^{3/2}(t)\biggr] \\
				&\quad+9x(1-x)\biggl[b_{1\sigma}^{\eta_{q(s)}}\ln(x)+b_{2\sigma}^{\eta_{q(s)}}\ln(1-x)\biggr],\\
			\end{split}
		\end{equation}
where $t=2x-1$ and $C$ functions are Gegenbauer polynomials. The coefficients in distribution amplitudes are as follows,
		\begin{equation}
			\begin{split}
				&a_1^{\eta_q}=0, \quad a_2^{\eta_q}=0.25\pm 0.15, \\
				&a_{0P}^{\eta_q}=0.079\pm 0.028, \quad a_{2P}^{\eta_q}=0.95\pm 0.33,\\
				&a_{4P}^{\eta_q}=0.14\pm 0.11, \quad a_{1P}^{\eta_q}=a_{3P}^{\eta_q}=0, \\
				&b_{1P}^{\eta_q}=b_{2P}^{\eta_q}=0.039\pm 0.014, \\
				&a_{0\sigma}^{\eta_q}=0.055\pm 0.024,\quad a_{2\sigma}^{\eta_q}=0.18\pm 0.07, \\
				&a_{1\sigma}^{\eta_q}=a_{3\sigma}^{\eta_q}=0, \quad b_{1\sigma}^{\eta_q}=b_{2\sigma}^{\eta_q}=0.026\pm 0.009, \\
			\end{split}
		\end{equation}
for the $\eta_q$ meson, and
		\begin{equation}
			\begin{split}
				&a_1^{\eta_s}=0, \quad a_2^{\eta_s}=0.25\pm 0.15, \\
				&a_{0P}^{\eta_s}=1.13\pm 0.41, \quad a_{2P}^{\eta_s}=0.99\pm 0.48,\\
				&a_{4P}^{\eta_s}=0.06\pm 0.05, \quad a_{1P}^{\eta_s}=a_{3P}^{\eta_s}=0, \\
				&b_{1P}^{\eta_s}=b_{2P}^{\eta_s}=0.56\pm 0.20, \\
				&a_{0\sigma}^{\eta_s}=0.79\pm 0.34,\quad a_{2\sigma}^{\eta_s}=0.14\pm 0.07, \\
				&a_{1\sigma}^{\eta_s}=a_{3\sigma}^{\eta_s}=0, \quad b_{1\sigma}^{\eta_s}=b_{2\sigma}^{\eta_s}=0.38\pm 0.14, \\
			\end{split}
		\end{equation}
for the $\eta_s$ meson.
		
For $ \rho $ meson, the twist-2 and twist-3 distribution amplitudes are given by \cite{PB-GWJ2007}
		\begin{equation}
			\phi_{\rho}(x)=6x(1-x)[1+a_{1}^{||} C_{1}^{3/2}(t)+a_{2}^{||}C_{2}^{3/2}(t)],
		\end{equation}
		\begin{equation}
			\begin{split}
				\phi_{\rho}^{t}(x)=&3t^{2}+ \dfrac{3}{2}a_{1}^{\perp}t(3t^{2}-1)+\dfrac{3}{2}a_{2}^{\perp}t^{2}(5t^{2}-3)+(\dfrac{15}{2}\kappa_{3}^{\perp}-\dfrac{3}{4}\lambda_{3}^{\perp})t(5t^{2}-3)+\dfrac{5}{8}\omega_{3}^{\perp}(35t^{4}-30t^{2}+3)\\
				&+\dfrac{3}{2}\dfrac{m_{q}+m_{q}}{m_{\rho}}\dfrac{f_{\rho}^{||}}{f_{\rho}^{\perp}}\biggl\{1+ 8a_{1}^{||}t+ 3a_{2}^{||}[7-30x(1-x)]+t(1+3a_{1}^{||}+6a_{2}^{||})\ln(1-x)\\&
				-t(1-3a_{1}^{||}+6a_{2}^{||})\ln x\biggr\}, \\
			\end{split}
		\end{equation}
		\begin{equation}
			\begin{split}
				\Psi_{3}^{||}(x)=&6x(1-x)\biggl[ 1+(\dfrac{1}{3}a_{1}^{\perp}+\dfrac{5}{3}\kappa_{3}^{\perp})C_{1}^{3/2}(t)+(\dfrac{1}{6}a_{2}^{\perp}+\dfrac{5}{18}\omega_{3}^{\perp})C_{2}^{3/2}(t) -\dfrac{1}{20}\lambda_{3}^{\perp}C_{3}^{3/2}(t)\biggr]\\
				&+3\dfrac{m_{q}+m_{q}}{m_{\rho}}\dfrac{f_{\rho}^{||}}{f_{\rho}^{\perp}}\biggl\{x(1-x)[1+2a_{1}^{||}t+3a_{2}^{||}(7-5x(1-x))]+(1+3a_{1}^{||}+6a_{2}^{||})(1-x)\ln(1-x)\\
				&+(1-3a_{1}^{||}+6a_{2}^{||})x\ln x\biggr\},\\
			\end{split}
		\end{equation}
where $t=2x-1$, $C$ functions are Gegenbauer polynomials and $ \phi_{\rho}^{s}(x)=\dfrac{1}{2}\dfrac{d\Psi_{3}^{||}(x)}{dx} $. The coefficients in distribution amplitudes are as follows,
		\begin{equation}
			\begin{split}
				&a_{1}^{||}=0, \quad a_{2}^{||}=0.15\pm0.07, \\
				&a_{1}^{\perp}=0, \quad a_{2}^{\perp}=0.14\pm0.06, \\
				&\kappa_{3}^{\perp}=0, \quad \omega_{3}^{\perp}=0.55\pm0.25,\\
				&\lambda_{3}^{\perp}=0, \quad m_{q}=0.0056\pm0.0016
			\end{split}
		\end{equation}
		
The above parameters are all determined at the renormalization scale of $ \mu = 1.0~\mathrm{GeV}$. The Gegenbauer polynomials are given by
		\begin{equation}
			C_1^{1/2}(t)=t, \quad
			C_2^{1/2}(t)=\frac{1}{2}\left(3t^2-1\right), \quad
			C_3^{1/2}(t)=\frac{1}{2}t\left(5t^2-3\right), \quad
			C_4^{1/2}(t)=\frac{1}{8}\left(35t^4-30t^2+3\right), \quad
		\end{equation}
and
		\begin{equation}
			C_1^{3/2}(t)=3t, \quad
			C_2^{3/2}(t)=\frac{3}{2}\left(5t^2-1\right), \quad
			C_3^{3/2}(t)=\frac{5}{2}t\left(7t^2-3\right), \quad
			C_4^{3/2}(t)=\frac{15}{8}\left(21t^4-14t^2+1\right). \quad
		\end{equation}
	\end{widetext}

\end{document}